\documentclass[10pt,final,twocolumn,twoside]{IEEEtran}
 \usepackage{setspace}
 \usepackage{calligra}
 \usepackage{latexsym}
 \usepackage{times}
 \usepackage{graphicx}
 \usepackage{color}
\usepackage{algorithm}
\usepackage{algorithmic}
 \usepackage{amssymb,amsmath}
 \usepackage{multirow}
 \usepackage{colortbl} 
\usepackage{subcaption}
\usepackage{enumerate}
\usepackage{graphicx}
\usepackage{caption}
\usepackage{mathrsfs}
\usepackage{amsbsy}
\usepackage{xcolor}
\usepackage{mathtools}
\usepackage[utf8]{inputenc}

\newcommand{\mbf}[1]{\mathbf{#1}}

\begin{document}
\title{CLIMEX: A Wireless Physical Layer Security Protocol Based on Clocked Impulse Exchanges}

%\author{ }
 \author{Satyam Dwivedi, John Olof Nilsson, Panos  Papadimitratos,
   Peter H\"{a}ndel} 
\maketitle

 % clksec
 \begin{abstract} 
A novel method and protocol establishing common secrecy   based
   on physical parameters between two users is proposed. The four physical parameters of users
   are their clock frequencies, their relative clock phases and the
   distance between them.  The protocol proposed between two users is backed by  theoretical model for the measurements. Further, estimators are proposed to estimate secret physical parameters. Physically exchanged parameters are shown
   to be secure by virtue of their non-observability to adversaries. Under a simplified analysis based on a testbed settings, it is shown that 
   $38$ bits of common secrecy can be derived for one run of the proposed protocol
  among users. The method proposed is also robust against various
  kinds of active timing attacks and active impersonating adversaries. 
 \end{abstract}

%\settopmatter{printfolios=true}

 \section{Introduction}
 % Physical properties are also increasingly finding their usage in
 % providing secrecy to information exchange between physically
 % separated entities. Usage of clocks as physical unclonable
 % functions (PUF) and usage of wireless channel in generating secret
 % keys are examples of physical properties providing authentication
 % of entities and encryption of information among entities
 % \cite{suh2007physical, PUF2, 5999759, bloch2011physical}.

  Clocks are everywhere. They are found in every gadget around us. Various operations across the
 digital system are synchronized at clock edges. Crystal clocks
 are used in most consumer electronics 
 present in our surroundings as heart beats of electronic
 systems. This motivates usage of physical characteristics of clocks
 in security applications and has the potential to provide security to far reaches of electronic presence. 

\begin{figure}
   \centering
   \includegraphics[height=2.5in]{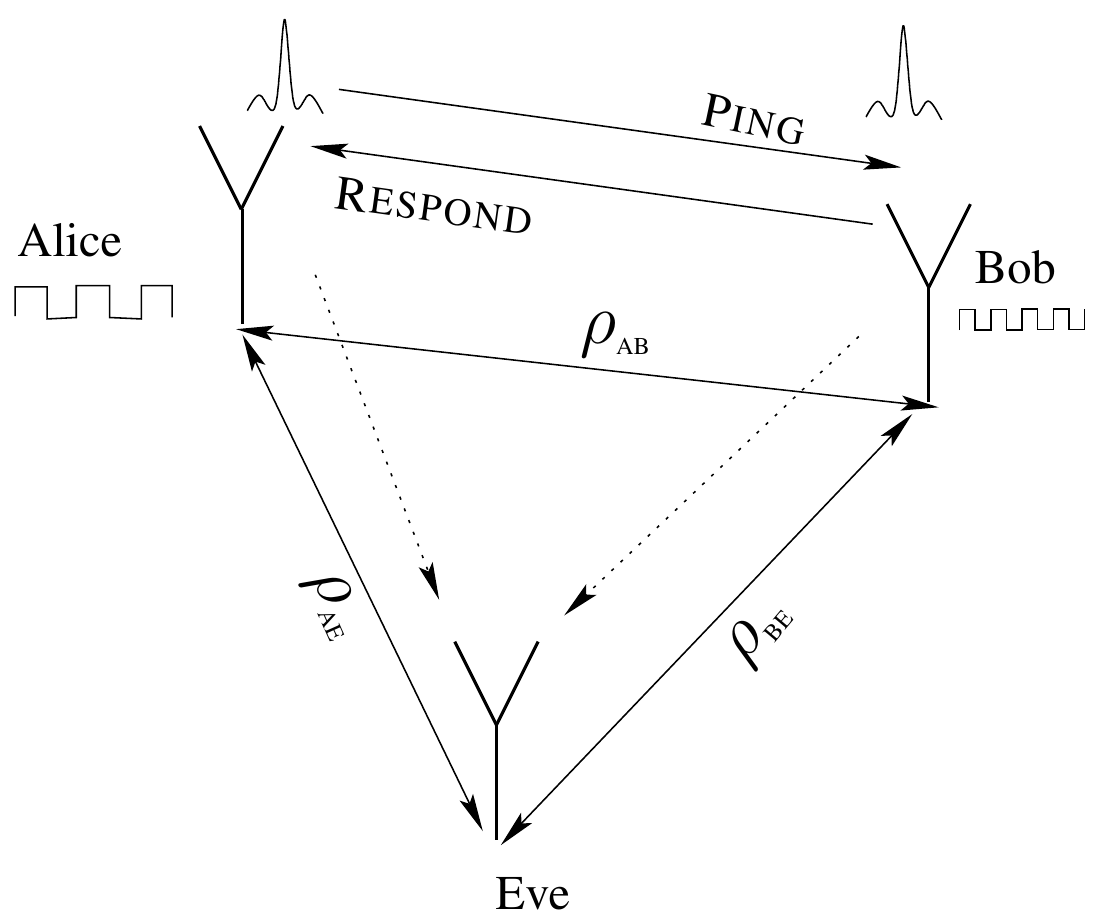}
   \caption{Alice, Bob and Eve setup. Alice and Bob separated by distance $\rho_{\text{\tiny AB}}$ exchange
     physical parameters over wireless to generate a shared secret
     key. Eve who is $\rho_{\text{\tiny AE}}$ distance away from Alice and $\rho_{\text{\tiny BE}}$ away from Bob, is an adversary listening to the wireless exchanges. She
     is attempting to decipher the parameters from
     measurements.}
   \label{fig:ABE}
 \end{figure}

Ever present electronics with clocks is synonymous to internet-of-things (IOT) context. With increasing potential of IOT applications, security of data exchange in IOT applications also is of paramount importance \cite{Weber201023, jing2014security}. The IOT context puts more stringent constraints on resource consumption for achieving
 secrecy. Low complexity
 schemes for key generation is proposed in \cite{amitav_IOT} As suggested in 
 \cite{phy_sec_challenges}, physical layer security can play an
 important role in IOT applications by alleviating the need for high
 computation required in traditional cryptography by making use of existing radio
 functionality to support security.  In \cite{prk_security_CPS},
 involvement of control and physical layer of a system for security of
 cyber physical system (CPS) is suggested.  In physical layer
 security,
 encrypting messages using secret keys
 generated based on wireless channel is the most investigated method
 \cite{phy_sec_challenges, 5999759, bloch2011physical, 7270404, phy_keygen_challenges,
    keygen_review, 6739367, physec_time_varying}.
 In an Alice, Bob and Eve setup as shown in Fig.\ref{fig:ABE}, secret
 key generation has been based on the uniqueness of channel between Alice
 and Bob than Alice-Eve or Bob-Eve channel. Most of above wireless based key
 generation methods rely on randomness of channel to produce a unique
 key between Alice and Bob. However, such schemes suffer from many
 weaknesses as discussed in \cite{phy_sec_challenges}. Assumptions
 such as Gaussian symmetric channel, dynamic requirement of the
 channel, weak adversary models etc. are shown to be weak on many
 instances. Many other practical implementation and usage issues in
 implementing wireless channel based physical layer security are
 discussed in \cite{keygen_review, phy_sec_challenges}.

A few other physical parameters are proposed for generating secret
 keys over wireless.  They are received signal strength (RSS),
 drifting oscillators, distance between nodes, radiometric signatures,
 quantum key generation etc \cite{noncrypt_survey, fingerprinting_survey, RSS_keygen, mahboob_gross, takesue2007quantum, secure_neighbour_Panos, radiometric_cite}. In
 \cite{noncrypt_survey}, exhaustive review
 on key generation from physical parameters are written. Various ways  of non-cryptographic
 authentication are described.
 In \cite{RSS_keygen}, effectiveness of RSS for secret key generation
 between two nodes is experimentally analysed and in absence of a highly
 dynamic environment, low entropy bits are generated for secret
 keys. The randomness of a physical clock parameter, i.e., its
 oscillating frequency has been suggested a few times
 \cite{clock_cfo_TCOM, clock_skew_TMC, MIMO_radiometric}.
 % In \cite{mahboob_gross}, authors have proposed to establish and
 % track relative clock parameters beween Alice and Bob setup.
 % Authors suggest, two clocks can uniformly and uniquely form a clock
 % frequency pairs within part-per-million (PPM) specifications of
 % clocks.
 A carrier frequency tracking framework and subsequent device
 authentication based on it is proposed in
 \cite{clock_cfo_TCOM}. Whereas, a multi-input multi-output (MIMO) device authentication based on
 carrier frequencies from all the transmitters is proposed in
 \cite{MIMO_radiometric}.  The papers suggesting oscillator properties
 for secure exchanges are mostly based on authenticating nodes by
 their signature oscillator characteristics. These papers mostly
 describe the system to the extent of proving uniqueness of absolute
 or relative oscillator characteristics, while having a weak or no
 adversary model assumptions. Security in clock synchronization
 schemes has also been discussed \cite{secure_clock_synch,
   secure_clock_synch2}. Usage of clock synchronization to detect
 man-in-the-middle attack is discussed in \cite{MITM_PRK}. Fundamental
 limits on security using clock synchronization is proposed along with
 clock synchronization protocol for security purposes. Among other contexts of
 physical layer security, physical layer security in sensor networks
 based on distributed detection have been recently discussed in
 \cite{amitav_IOT, distributed_sec1}.

 Usage of distance and position for securing communication has often
 been discussed \cite{noncrypt_survey, fingerprinting_survey,
   RSS_keygen, sec_posit_JSAC, rasmussen2010realization, Brands1994,
   salimi2016pairwise}. Usage of distance between nodes implicitly
 appears in RSS based authentication and secure key generation methods
 \cite{noncrypt_survey, fingerprinting_survey, RSS_keygen}.  Whereas,
 explicit application of distance for security appears in concepts
 such as distance bounding protocols and their use in concepts such as
 secure neighbourhod discovery \cite{secure_neighbour_Panos,
   rasmussen2010realization, Brands1994, secure_vehicle_Panos,
   tippenhauer2015uwb, neuberg2011mobile}.

Vulnerability to active adversaries has been a very critical issue in
secure communications and key exchanges \cite{mani_active_adv,
  liu2010securing, papadimitratos2006securing, 5601960, sec_posit_JSAC}. There is limited
discussion on active adversary in physical layer context as pointed
out in
\cite{mani_active_adv}. In this paper usage of the 
proposed method in warding off adversary attack is also
suggested. Various active adversary models are suggested  in 
\cite{sec_posit_JSAC} which are very relevant to this work and
considered in subsequent sections of the paper.   

 The main motivation of this paper is proposing a secure mechanism to  exchange physical parameters of users among them over wireless.   Accordingly, the main idea proposed in this paper is a protocol which helps in secret exchange of physical parameters between two nodes. The proposed protocol and the measurements based on the protocol lets two nodes exchange their relative physical parameters. The protocol incorporates actions at two nodes to strengthen the robustness of parameter exchange against any passive eavesdropper and to some extent to an active adversary, as explained later in the paper. The  protocol exploits a physical phenomena between two continuous running clocks separated by a distance for parameter exchange while providing secrecy.

Any
 two physical clocks can have a frequency and a phase relation. The frequency relation considered here is the difference in frequency
 ($f_d$) between the clocks and the phase relation ($\phi$) is the relative
 phase of the clocks. If these clocks are mounted on two
 separate devices then the distance between these user ($\rho$)
 provides another physical dimension to use.  Further in the
 paper,  a set of
 four independent 
 parameters i.e., two clock frequencies of users derived from
 difference frequencies, their relative phase and the
 distance between them are shown to be exchanged as shared secret
 parameters between two users.  The idea presented
 in this paper is outcome of 
 experiments with impulse radio ultrawideband (IR-UWB) testbed
 \cite{UWB_testbed, satyam_clk1} that manifests practical relevance of
 the work. 

 In this paper IR-UWB signals are suggested to be used for signalling
  among Alice and Bob  in Fig.\ref{fig:ABE}.  Alice periodically sends \textsc{Ping} signals to Bob and after a delay Bob sends a \textsc{Respond} signal. The nodes Alice and Bob
 exchange impulse signals \textsc{Ping} and \textsc{Respond} triggered by
 time events generated using an oscillator or a clock. Hence
 the mechanism or the protocol of signal exchange will be called as CLocked IMpulse EXchange
 (CLIMEX). Further discussions in the paper is based on classical
 Alice, Bob and Eve setup \cite{bloch2011physical, 7270404}.

The main contributions of the paper are:
 \begin{enumerate}
 \item The CLIMEX protocol between two nodes to exchange physical parameters of two nodes
   discreetly over wireless. 
 \item A mathematical model for the measurements at nodes. 
 \item Estimators for estimating required parameters at nodes from the
   collected measurements.
 \item A simplified analysis to evaluate number of possible secret bits which can be derived
   from epochs of measurements.  Secret bits derived can
   be further used for encryption and authentication. 
 \item Evaluation in presence of passive adversary. Different possible measurements by passive adversary are considered in this paper. 
  \item Discussing robustness against a few types of active
    adversaries. Particularly, against timing attacks and 
   impersonating adversary. 
 \end{enumerate}
 \begin{figure}
  \centering
   \includegraphics[width=3.5in]{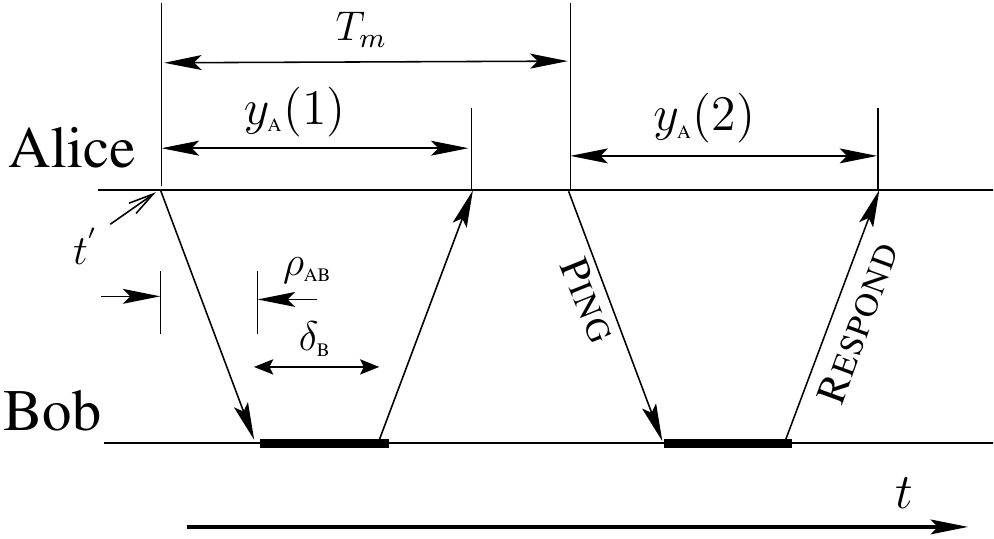}
   \caption{Conceptual timing diagram of one-way RTT protocol between
    Alice and Bob. Alice transmits \textsc{Ping} at interval $T_m$ seconds and collects measurements $\left [y_{\text{\tiny A}}(1)\ y_{\text{\tiny A}}(2) \cdots \right
 ]^\top$. Bob is a distance $\rho_{\text{\tiny AB}}$ away from Alice and sends \textsc{Respond} after a delay $\delta_{\text{\tiny B}}$ from the instance it receives the \textsc{Ping} from Alice.}
   \label{1way}
 \end{figure}
 In summary, we propose analyse and provide experimental observations for CLIMEX as a tool for secure exchange of physical parameters. 

 This paper is organised as follows.  Section \ref{delaysec} describes the principle of operation, gives an overview of measurements, describes the basic phenomenon to be used and introduces the CLIMEX protocol while comparing it with the basic RTT protocol. Section \ref{sys_mod}
 describes the basic scheme and compares the proposed CLIMEX
 protocol with basic RTT protocol. Next described are measurement models
 for the protocols in section \ref{secmeas}. Estimators based on
 measurement models are presented in \ref{estim_AB}. Following estimators is the section \ref{adv_mod} on adversary models which discusses possible measurements adversary can do. 
  Section \ref{secanal} has analyzed the security aspects proposed in this paper.  It is followed by further discussion on practicality and possible future work on the proposed idea in section \ref{secdis}. The last section then is the conclusion. 

\begin{figure*}
   \centering
   \includegraphics[height=2in]{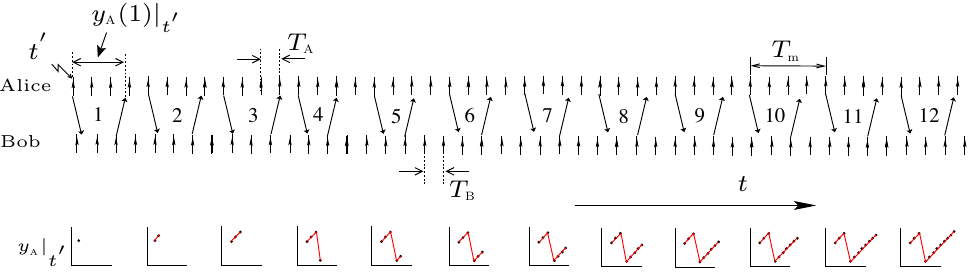}
   \caption{Interplay between two clocks giving rise to sawtooth
     measurements at the measuring end.}
   \label{fig:clkmech}
 \end{figure*}
 \section{Principle of Operation}
\label{delaysec}

The principle of parameter exchange among nodes proposed in this paper is based on round trip time (RTT) measurements.				
One-way RTT is shown in the
 Fig.\ref{1way}. Alice sends  '\textsc{Ping}' signals to Bob at regular interval of $T_m$ seconds. Bob 
 responds after a delay
 $\delta_{\text{\tiny B}}$ and does not collect any measurement. The delay $\delta_{\text{\tiny B}}$ is generated at Bob's end using its own clock with frequency $f_{\text{\tiny B}}$. Alice
 collects an epoch of RTT
 measurements,
 $\left [y_{\text{\tiny A}}(1)\ y_{\text{\tiny A}}(2) \cdots \right
 ]^\top$. These measurements are collected using a precise time difference measurement device which usually has time measurement precision much higher than clock periods of the clocks in discussion. For very clock frequencies (MHz or greater) in consideration, typically the precision is in range of nano to pico seconds. One such device is time to digital converter (TDC) \cite{acam_TDC, TI_TDC}.  Usage of such measurement precision is  also customary in network time synchronization schemes like precision time protocol (PTP) \cite{PTP_wiki, PTP2}. It is followed by Bob also collecting an epoch of measurements
 $\left [y_{\text{\tiny B}}(1)\ y_{\text{\tiny B}}(2) \cdots \right ]^\top$
 by sending several '\textsc{Ping}s' and measuring RTT where Alice introduces delay $\delta_{\text{\tiny A}}$ using its clock with frequency $f_{\text{\tiny B}}$. 
 \label{sys_mod}
 From their wireless signal
 exchanges and RTT measurements, Alice and Bob can estimate a few relative parameters among them.
 They can estimate their relative clock frequency 
 $f_d = f_{\text{\tiny A}} - f_{\text{\tiny B}}$, relative
 phase of their clocks, $\phi_{\text{test}}$ and the parameter
 $\rho_{\text{\tiny AB}}$, the distance between Alice and Bob
 \cite{UWB_testbed, satyam_clk1}. The first measurements at
 Alice in the sequence of measurements is timestampped with time
 $t^{'}$. The phase of the signal in the measurement depends on the instance of the measurement collection.
It is assumed
 that the physical parameters to be exchanged remain time invariant
 during this period when Alice and Bob collect their respective
 measurements. 

 \subsection{Interplay between two clocks}

As shown in Fig.\ref{fig:clkmech}, Alice and Bob periodically exchange
\textsc{Pings} and corresponding \textsc{Responds}. Figure
\ref{fig:clkmech} illustrates generation of sawtooth waveform as a
result of interplay between clocks of Alice and Bob through RTT
measurements among them.  Alice's clock has period $T_{\text{\tiny A}} =
1/f_{\text{\tiny A}}$ and Bob's clock has period $T_{\text{\tiny B}} =
1/f_{\text{\tiny B}}$. Alice sends successive
\textsc{Pings} at regular interval, $T_{\text{\tiny m}} = 4
T_{\text{\tiny A}}$ in the illustration. Bob \textsc{Responds} after
introducing a delay of one clock period $T_{\text{\tiny B}}$ from the clock edge subsequent
to arrival of signal from
Alice. As illustrated form the figure, at instance 3, Bob receives
\textsc{Ping} from Alice just after a clock edge has elapsed, which
results in largest RTT measurement at Alice. Whereas, at instance 4,
Bob receives the \textsc{Ping} nearly at the same time a clock edge
arrives. Hence, the RTT at Alice is the smallest. The frequency of sawtooth waveform is the
difference frequency $f_d$. The height of the sawtooth measurement equals $T_{\text{\tiny B}}$. The phase of both clocks are related to
the phase of the sawtooth waveform which is a function of $t^{'}$, instance when the measurement begins. Generation of sawtooth from RTT measurements
was first reported in 
\cite{satyam_clk1}. The above protocol of signal exchange between Alice and Bob is RTT protocol. CLIMEX protocol
with a few changes in RTT protocol is introduced next. 
 \begin{figure}
   \begin{subfigure}{0.5\textwidth}
     %\centering
     \includegraphics[width=0.9\textwidth]{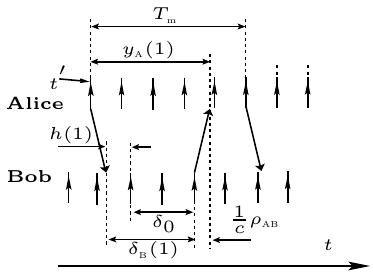}
     \caption{The RTT protocol.}
     \label{detailed_RTT1}
   \end{subfigure}
   \vfill
   \begin{subfigure}{0.5\textwidth}
  %   \hspace{0.3in}
     \includegraphics[width=0.9\textwidth]{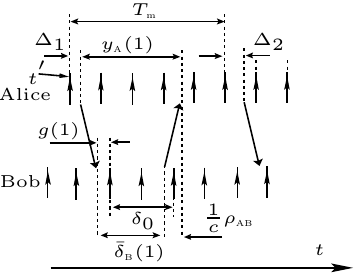}
     \caption{The CLIMEX Protocol.}
     \label{detailed_RTT}
   \end{subfigure}
   \hspace{-0.3in}
   \caption{The RTT protocol and the CLIMEX protocol. In addition to the RTT protocol, the CLIMEX protocol has random
     transmission delays $\Delta_1$, $\Delta_2$ introduced by Alice and scaled
     delay $\bar\delta_{\text{\tiny B}}(1)$ at Bob's end.}
\label{RTTprots}
 \end{figure}
 \subsection{RTT protocol and the CLIMEX protocol}
 \label{protocols}  RTT measurements where
 events are initiated by clock edges at both the ends are shown in
 Fig.\ref{RTTprots}. 
  The figure has clock tick level detail of RTT measurements shown 
 in Fig.\ref{1way}. Figure \ref{detailed_RTT1} is the RTT
 protocol resulting in sawtooth RTT measurements and Fig.\ref{detailed_RTT} is the CLIMEX
 protocol. RTT and CLIMEX protocols are explained now.
 \begin{enumerate}
 \item The upside arrows in the Figs. \ref{detailed_RTT1} and
   \ref{detailed_RTT} are the clock edges(ticks) of Alice's and Bob's
   clocks. As in the Fig.\ref{detailed_RTT1}, \textsc{Ping} and
   \textsc{Respond} events are generated on clock edges.
 \item Alice transmits \textsc{Ping} signals periodically every $T_m$
   seconds.  Alice counts this period using its clock with 
   clock period $T_{\text{\tiny A}}$. In the CLIMEX protocol of Fig. \ref{detailed_RTT}, instead of transmitting \textsc{Ping} at clock edge, 
   Alice transmits after a random delay $\Delta_1$ from the clock
   edge.    The added random delay is known only to the Alice. It provides secure
   exchange of keys by dithering RTT measurements as 
   explained later in \ref{secclimex}.

 \item Bob receives the \textsc{Ping} after a path delay of
   $\rho_{\text{\tiny AB}}/c$ seconds, $c$ is speed of light. In
   the RTT protocol of Fig.\ref{detailed_RTT1}, Bob responds to
   Alice's ping after a nominal wait, $\delta_0$, of two clock
   periods. The total delay at Bob's end would be
   $\delta_{\text{\tiny B}}(1)$. In the CLIMEX
   protocol, Bob responds to Alice's \textsc{Ping} after a scaled
   delay of $\bar\delta_{\text{\tiny B}}(1)$ seconds as shown in
   Fig.\ref{detailed_RTT}. The scaled delay
   $\bar\delta_{\text{\tiny B}}$ will be explained in \ref{secclimex}.
 \item Alice receives Bob's \textsc{Respond} and measures RTT
   $y_{\text{\tiny A}}(1)$ as shown in Figs. \ref{detailed_RTT1} and
   \ref{detailed_RTT}.
 \item Alice collects an epoch of $N$ RTT measurements. Measurements are
   timestampped by time measurement beginning, $t'$.
   $\mbf{y_{\text{\tiny A}}}|_{t'} = [y_{\text{\tiny A}}(1) \cdots \:
   y_{\text{\tiny A}}(N)]^\top \in \mathbb{R}^N$.  
 \item After Alice, Bob collects his measurements in a
   similar way as Alice. Bob's epoch of measurements is
   $\mbf{y_{\text{\tiny B}}}|_{t''} = [y_{\text{\tiny B}}(1) \cdots \:
   y_{\text{\tiny B}}(N)]^\top \in \mathbb{R}^N$ which is timestampped
   by $t''$.
 \end{enumerate}
 As evident from the above discussion that the main differences
 between the RTT protocol and the CLIMEX protocol are addition of
 random delay $\Delta_i$ to \textsc{Ping}s by Alice and the scaled
 \textsc{Respond} $\bar\delta_{\text{\tiny B}}$ by Bob. In the next
 section it will be shown that the proposed changes to the CLIMEX
 protocol over RTT protocol gives rise to a  measurement
 model which helps in establishing common secrecy between Alice and Bob.
 % \subsection{The measurement and the noise models}

 \section{Measurement Model} \label{secmeas} In this section, 
 measurement model of RTT protocol as shown in Fig.\ref{detailed_RTT1}
 and CLIMEX protocol as shown in Fig.\ref{detailed_RTT} is suggested. As
 subsequently shown, both the measurement model is nonlinear in parameters
 of interest and model of CLIMEX protocol provides useful features for secure
 exchanges of parameters.

 \subsection{Measurement model for an epoch collection by Alice
   one-way in RTT protocol} 
 RTT measurement epoch recorded by Alice is 
 $\mbf{y_{\text{\tiny A}}}|_{t'} = [y_{\text{\tiny A}}(1) \cdots \:
 y_{\text{\tiny A}}(N)]^\top \in \mathbb{R}^N$.  It can be written compactly in vector form
 as \begin{eqnarray}\label{eq:m11} \mbf{y_{\text{\tiny A}}}|_{t'} =
   \boldsymbol{\delta}_{\text{B}} + \frac{2\rho_{\text{\tiny AB}}}{c}
   \mbf{1} + \mbf{w}\, ,
    \end{eqnarray}    where $\boldsymbol{\delta}_{\text{B}} $ is the total delay vector
    at Bob's node
    \begin{eqnarray}\label{eq:m22} 
      \boldsymbol{\delta}_{\text{B}}&=&  \mbf{h}(f_d,  T_{\text{\tiny B}}, \phi^{'}, \mbf{n})  + \delta_0 \mbf{1}. 
    \end{eqnarray}
    The function
    $\mbf{h}(f_d, T_{\text{\tiny B}}, \phi^{'}, \mbf{n}) = [h(1)
    \cdots \: h(N)]^\top$ results in sawtooth shape in measurements
    and can be given by
    % \subsubsection{
    %   $\mbf{h}(f_d, T_{\text{\tiny B}}, \phi^{'}, \mbf{n})$ in RTT
    %   protocol measurements} 
    \begin{eqnarray}\label{eq:m33}
      \mbf{h}(f_d,  T_{\text{\tiny B}}, \phi^{'}, \mbf{n}) \triangleq 
      \text{mod}_{T_{\text{\tiny
      B}}} \Bigg( \!\frac{\displaystyle T_{\text{\tiny B}}}{\displaystyle 2 \pi} \text{mod}_{2
      \pi}(2\pi f_d\mbf{t} + \phi^{'} \mbf{1}) + \mbf{n}\! \Bigg). 
    \end{eqnarray}
    The time vector
    $\mbf{t} = [t' \ t'+T_m \ t'+2T_m\ \cdots \
    t'+(N-1)T_m]^\top$.
%    $f_{\text{\tiny A}} \in [f_{\text{\tiny A}}^{\text{\tiny min}} \
%   f_{\text{\tiny A}}^{\text{\tiny max}}]$.  Similarly,
%   $ f_{\text{\tiny B}} \in [f_{\text{\tiny B}}^{\text{\tiny min}} \
%    f_{\text{\tiny B}}^{\text{\tiny max}}]$. 
$\phi^{'}$ is the
    relative phase of Alice's and Bob's clock when the measurement has
    timestamp $t'$ with first measurement $y_{\text{\tiny A}}(1)
    $. Similarly, $\phi^{''}$ is the phase of sawtooth where the
    measurement epoch is collected by Bob. The noise vectors $\mbf{n}$
    and $\mbf{w}$ are explained in detail in the subsection \ref{noisesec}.

\begin{figure}
      \begin{subfigure}{0.5\textwidth}
        \centering \hspace{-0.5in}
        \includegraphics[width=1\textwidth]{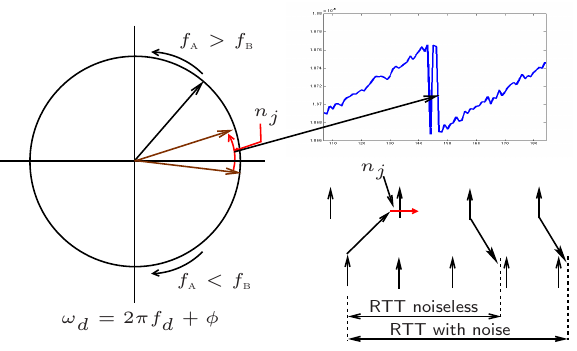}
        \caption{Phasor for sawtooth measurements. Shows noise effect
          in jumping quadrant leading to zero crossings.}
        \label{phasor}
      \end{subfigure}
      \begin{subfigure}{0.4\textwidth}
        \centering
        \includegraphics[width=1\textwidth]{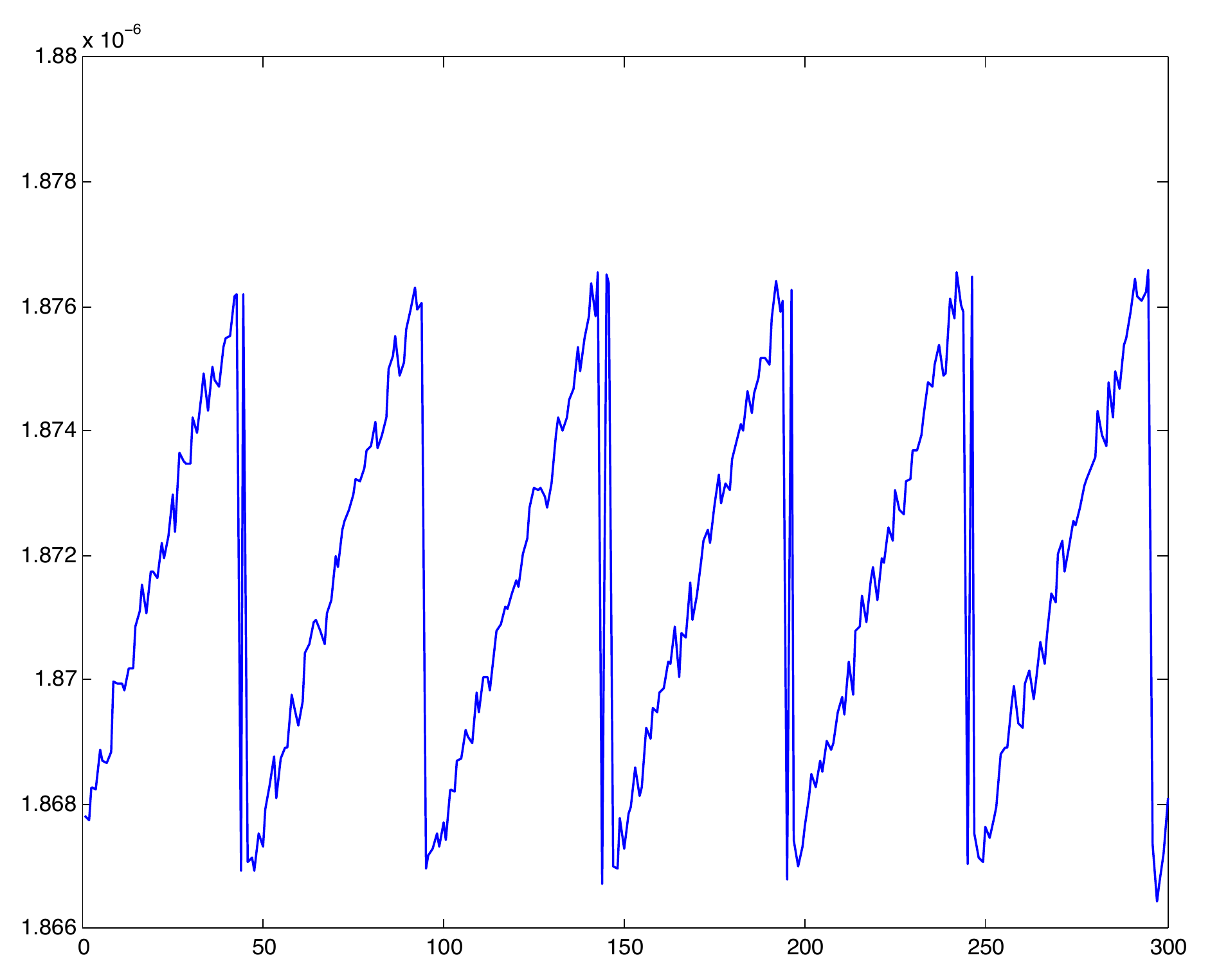}
        \caption{Measurements at Alice/Bob, Sawtooth has positive
          slope as $f_{\text{\tiny A}}$ is larger than $f_{\text{\tiny B}}$, for $f_d = f_{\text{\tiny A}} - f_{\text{\tiny B}}$.}
        \label{SP}
      \end{subfigure}
      \hspace{0.1in}
      \begin{subfigure}{0.4\textwidth}
        \centering
        \includegraphics[width=1\textwidth]{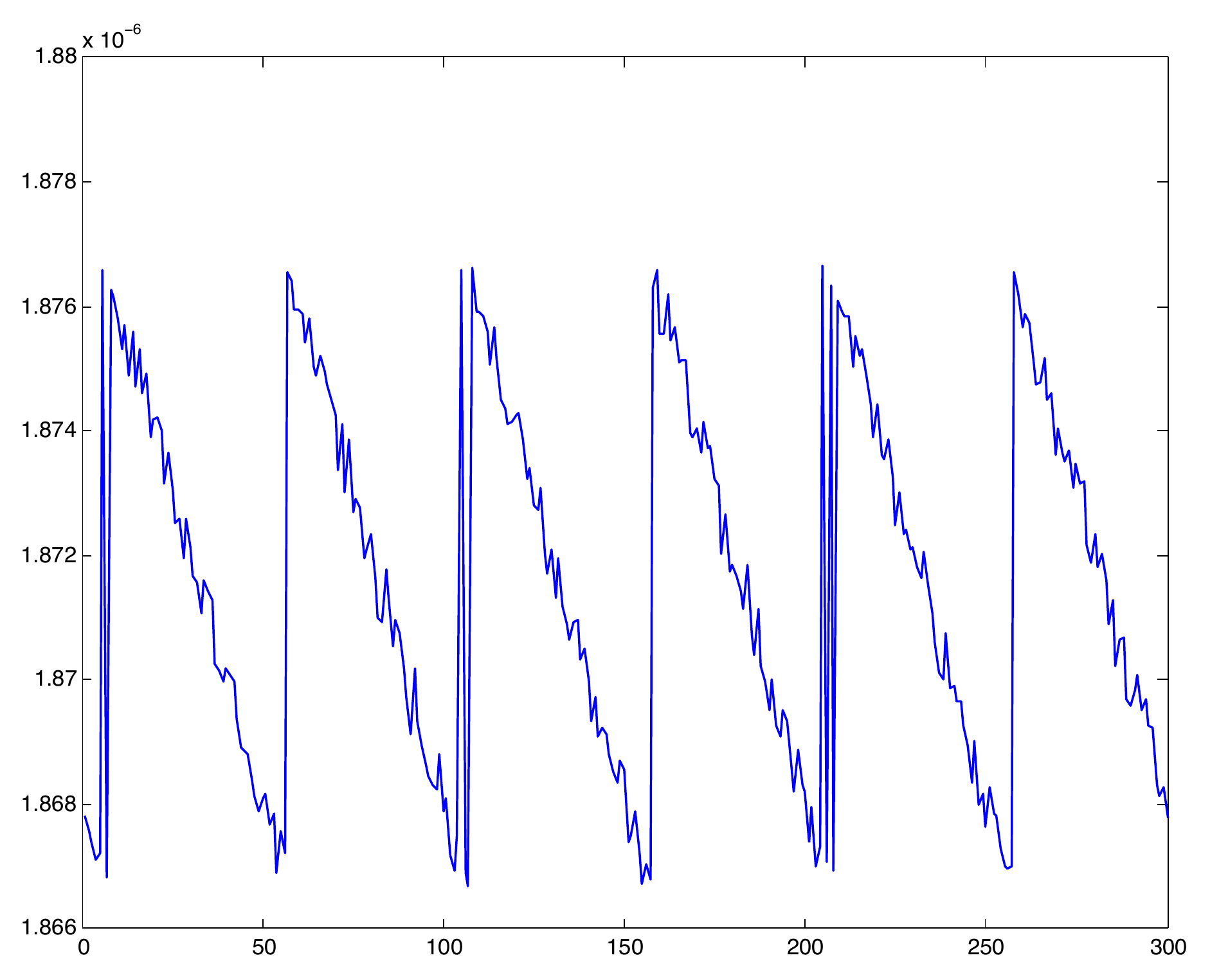}
        \caption{Measurements at Bob/Alice, Sawtooth has negative
          slope as $f_{\text{\tiny A}}$ is smaller than $f_{\text{\tiny B}}$, for $f_d = f_{\text{\tiny A}} - f_{\text{\tiny B}}$.}
        \label{SN}
      \end{subfigure}
      \caption{Details of measurement protocol, phasor argument of the
        sawtooth function and effect of noise. }
      \label{sawtooths}
    \end{figure}

    There are two modulus terms in (\ref{hdef2}). The internal modulus
    is with respect to $2\pi$ radians which corresponds to one period
    of the sawtooth waveform \cite{satyam_clk1}. The outer modulus is
    taken with respect to $T_{\text{\tiny B}}$. The outer modulus was
    not considered  before as $\sigma^2 _j$ and
    $\sigma^2 _c$ were assumed small \cite{satyam_clk1}. The model with only internal
    modulus was accurate enough to estimate parameters of interest
    with sufficient accuracy from the experimental data. However, in
    the current paper, random delays ($\Delta_i$) are added to the
    argument of the outer modulus. Hence the outer modulus comes
    into picture. Explanation for outer modulus is seen from Fig.\ref{phasor} which
    is a phasor diagram for argument of the sawtooth measurement. A
    noise term $n_j$ can move a phasor from one quadrant to
    another. Such transitions of moving a signal to the first quadrant
    results in nonlinear jumps in sawtooth waveform at
    discontinuities, as shown in Fig.\ref{phasor}. Also shown in the figure are effect of the phasor jump on RTT measurements. '\textit{RTT noiseless}' is RTT without any noise. Whereas, '\textit{RTT with noise}' is the RTT measurement when noise appears at phase discontinuity of sawtooth waveform. The sawtooth
    waveforms at Alice for two cases depending on the sign of $f_d$ is
    shown in Fig.\ref{SP} and Fig.\ref{SN}. In Fig.\ref{SP}, Alice's
    clock frequency is larger than Bob and in \ref{SN}, it is smaller.  The characteristic sawtooth waveform
    provides a valid signature to message exchange between Alice and
    Bob as discussed later.
 \begin{figure*}
      \centering
      \begin{subfigure}{0.195\textwidth}
        \centering
        \includegraphics[width=1\textwidth]{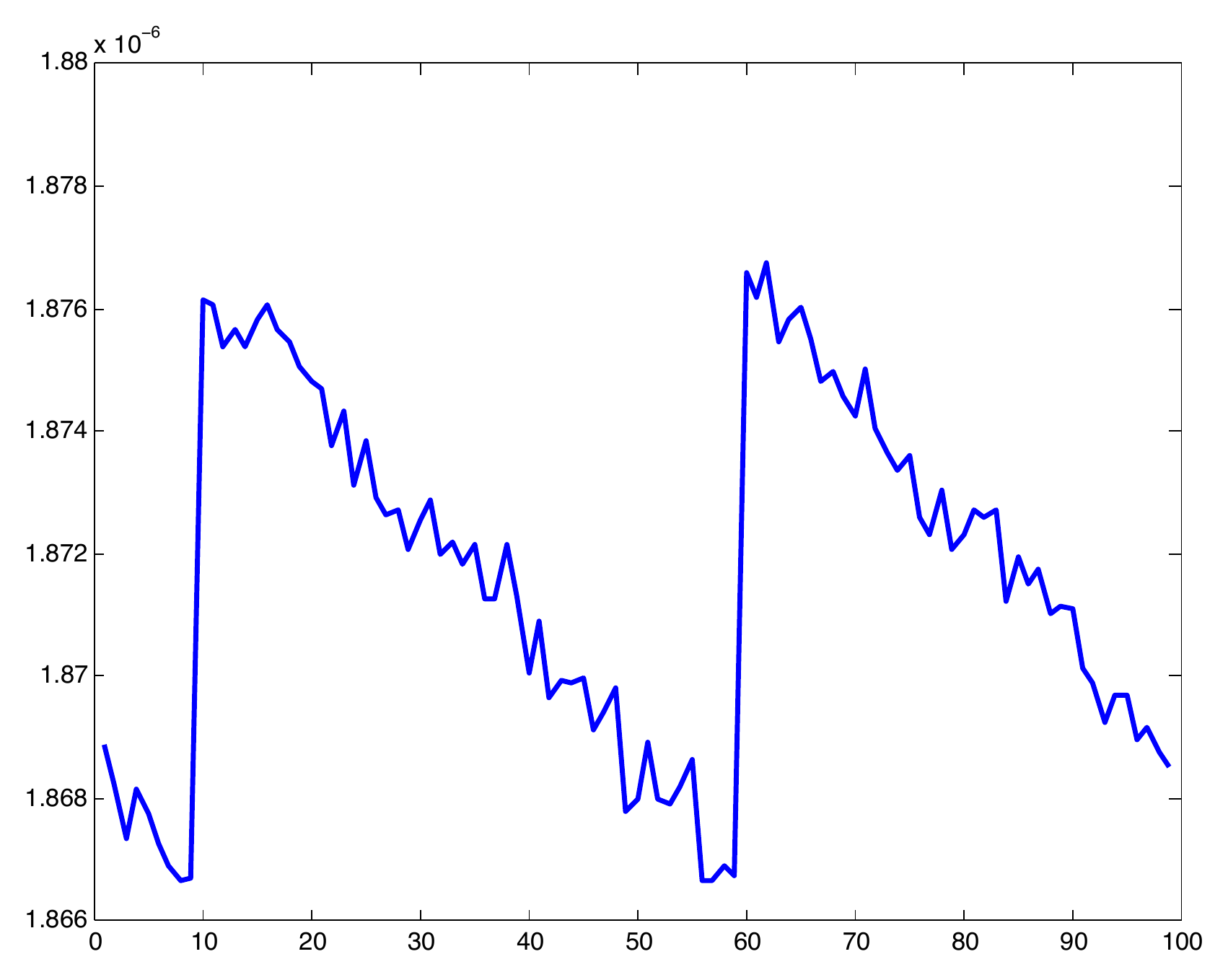}
        \caption{$\, \Delta \sim \mathcal{U}(0, 0)$}
        \label{dither1}
      \end{subfigure}
      \hfill
      \begin{subfigure}{0.195\textwidth}
        \centering
        \includegraphics[width=1\textwidth]{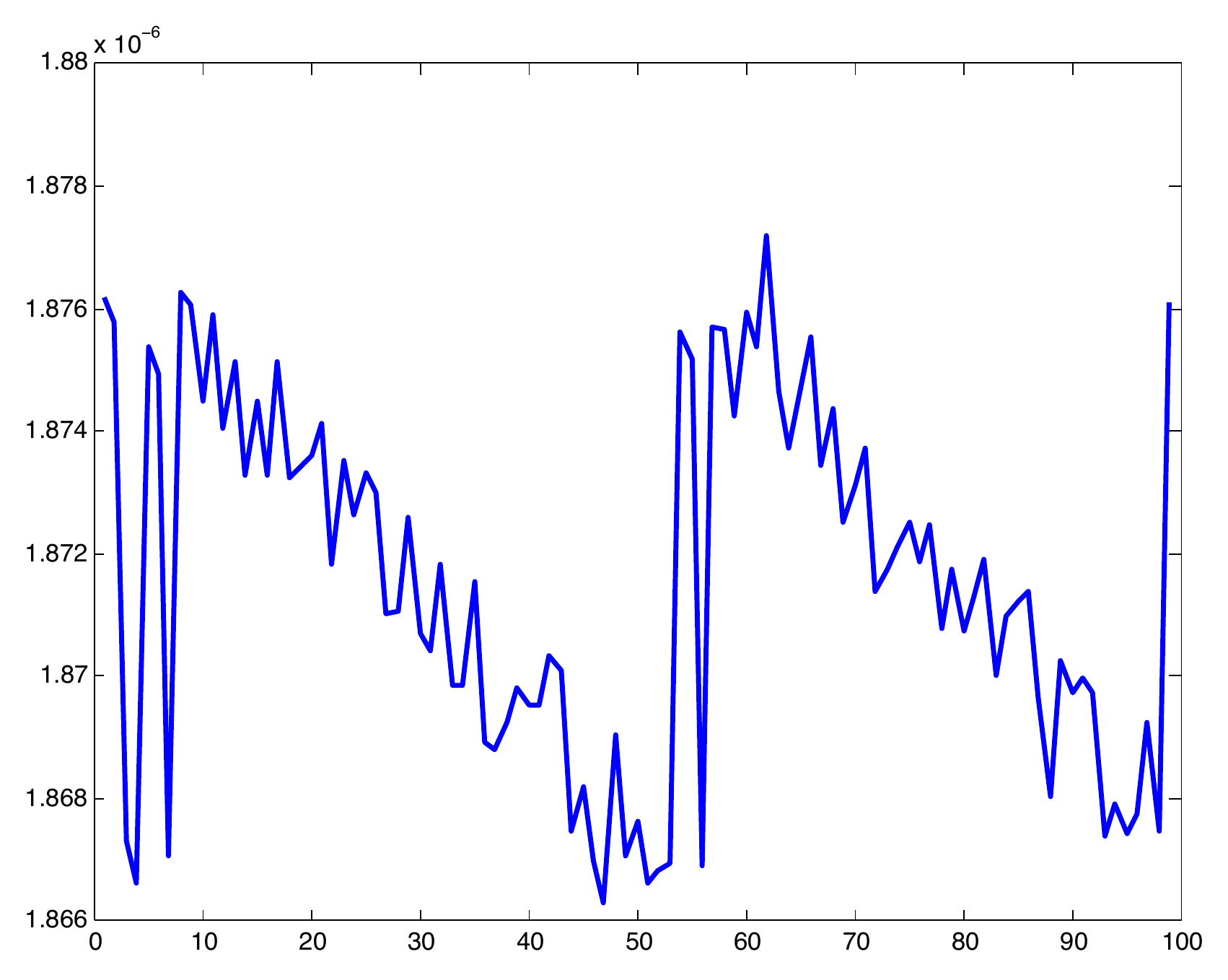}
        \caption{$\, \Delta \sim \mathcal{U}(0, 0.25 T_{\text{\tiny A}})$}
        \label{dither2}
      \end{subfigure}
      \hfill
      \begin{subfigure}{0.195\textwidth}
        \centering
        \includegraphics[width=1\textwidth]{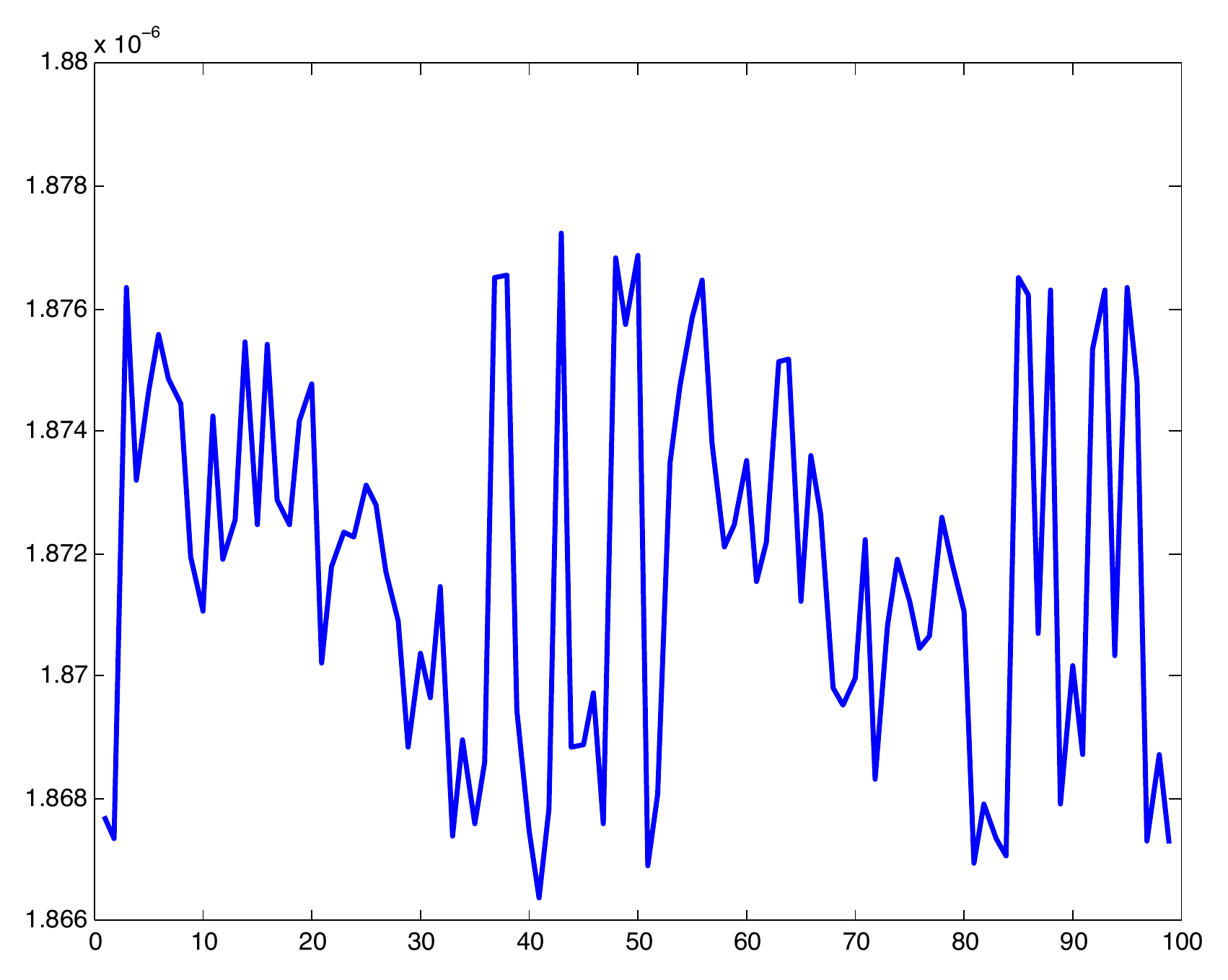}
        \caption{$\, \Delta \sim \mathcal{U}(0, 0.5 T_{\text{\tiny A}})$}
        \label{dither3}
      \end{subfigure}
      \hfill
      \begin{subfigure}{0.195\textwidth}
        \centering
        \includegraphics[width=1\textwidth]{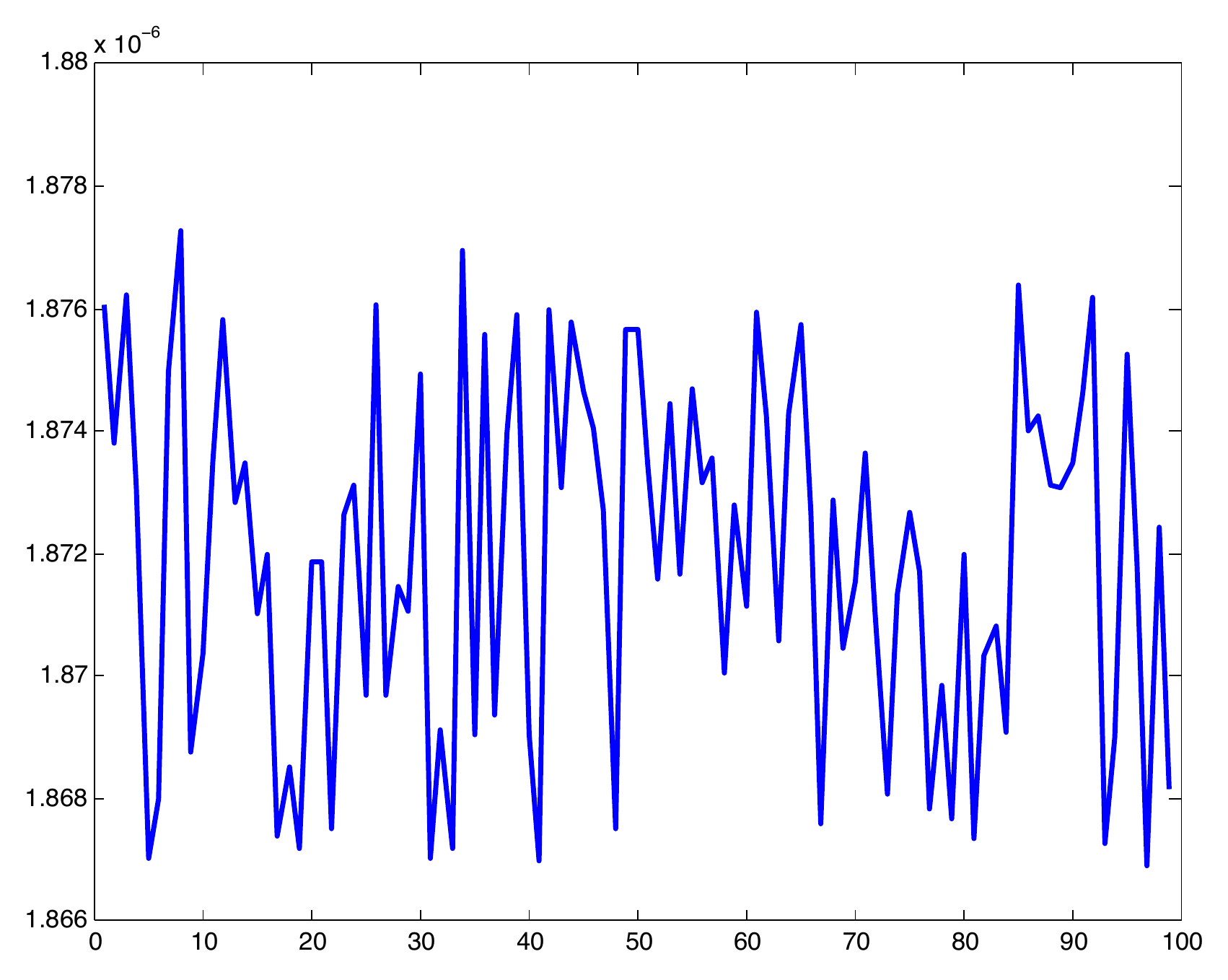}
        \caption{$\, \Delta \sim \mathcal{U}(0, 0.75 T_{\text{\tiny A}})$}
        \label{dither4}
      \end{subfigure}
      \hfill
      \begin{subfigure}{0.195\textwidth}
        \centering
        \includegraphics[width=1\textwidth]{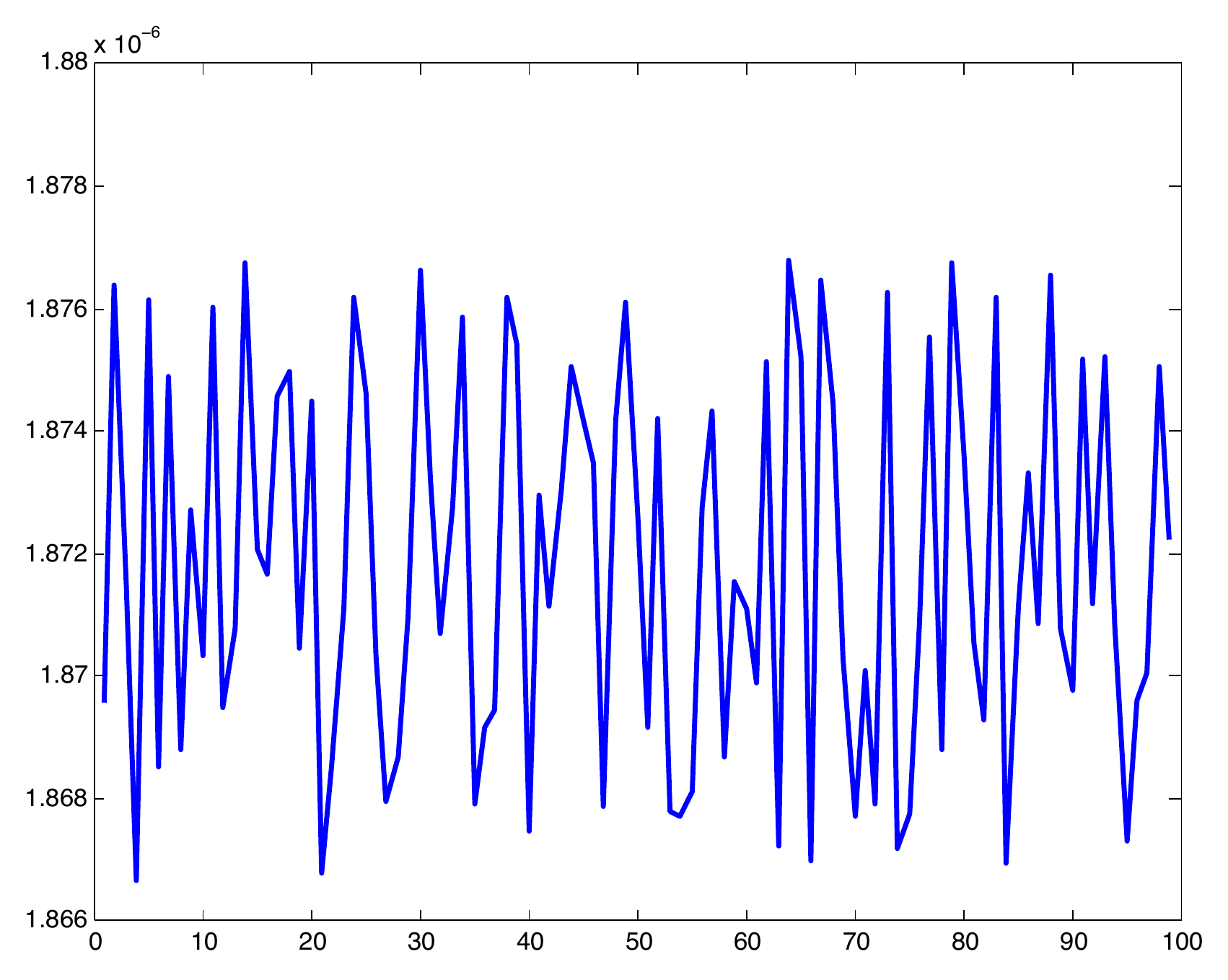}
        \caption{$\, \Delta \sim \mathcal{U}(0, T_{\text{\tiny A}})$}
        \label{dither5}
      \end{subfigure}
      \caption{RTT Measurements as a result of added random
        \textsc{Ping} delays $\mathbf{\Delta}$. As illustrated in the
        figure, increasing standard deviation of $\mathbf{\Delta}$ increases
        dithering of measurements. $\mathbf{\Delta}$ is known only to
        the node transmitting \textsc{Pings}. Any other node doing
        measurements will not be able to infer any information from
        it, due to lack of any information about $\mathbf{\Delta}$. }
      \label{ditherall}
\end{figure*}
    \subsection{Measurement model for an epoch collection by Alice in
      CLIMEX protocol} \label{secclimex}
    RTT measurements recorded by Alice after running CLIMEX
    protocol with Bob as shown in Fig.\ref{detailed_RTT} can be
    written
    \begin{eqnarray}\label{climeq1}
      \mbf{y_{\text{\tiny A}}}|_{t'} = \boldsymbol{\bar\delta}_{\text{B}} + \frac{2\rho_{\text{\tiny AB}}}{c}
      \mbf{1} + \mbf{w}\, ,
    \end{eqnarray}
    where $\boldsymbol{\bar\delta}_{\text{B}} $ is the total delay
    vector at Bob's node,
    \begin{eqnarray}
      \boldsymbol{\bar\delta}_{\text{B}}&=&  \mbf{g}(f_d, A, \phi^{'},
                                            \mathbf{\Delta}, \mbf{n})
                                            + \delta_0 \mbf{1}. 
                                            \label{bob_eq1}
    \end{eqnarray}
    The function
    $\mbf{g}(f_d, A, \phi^{'}, \mbf{n}) = [g(1) \cdots \: g(N)]^\top$
    results in sawtooth shape in measurements. The function
    $\mbf{g}(f_d, T_{\text{\tiny B}}, \phi^{'}, \mathbf{\Delta},
    \mbf{n})$ in this case is
    \begin{eqnarray}
      \mbf{g}(f_d, A, \phi^{'}, \mathbf{\Delta}, \mbf{n}) & \nonumber \\ \! \! \!
      \! \! \! \! \! \! \triangleq 
      \frac{A}{T_{\text{\tiny B}}}\text{mod}_{T_{\text{\tiny B}}}&\! \! \! \! \!\Bigg(\frac{\displaystyle T_{\text{\tiny B}}}{\displaystyle 2 \pi} \text{mod}_{2
                                                                   \pi}(2\pi f_d\mbf{t} + \phi^{'} \mbf{1}) + \mbf{\Delta} + \mbf{n}\Bigg)  \, . 
                                                                   \label{hdef2}
    \end{eqnarray}
    In (\ref{hdef2}), $\mbf{\Delta}$ is the random delay vector,
    $\mbf{\Delta} = [ \Delta_1, \Delta_2, ..., \Delta_N ]^\top$.  Delay
    $\Delta_1$  shown in Fig.\ref{detailed_RTT} is
    known only to Alice and can be assigned any distribution in order
    to increase estimation error of parameters at Eve. Figure \ref{ditherall} shows RTT measurements at
    Alice when random transmission delay uniformly distributed between
    is $0$ and $T_{\text{\tiny A}}$ are added while sending
    \textsc{Ping}s.  The figure shows the dithering effect delays
    $\mbf{\Delta}$ have on RTT measurements. Figure \ref{dither1}
    corresponds to measurements with 
    no random delay added to \textsc{Pings} and the measurement
    has the sawtooth waveform. In subsequent figures \ref{dither2}, \ref{dither3},
      \ref{dither4} and \ref{dither5}  increasing variance of the added random delay
    increasingly distorts the measurement waveforms.  Alice can estimate the shared secret 
    parameters  these measurements with the knowledge of
    $\mbf{\Delta}$ as will be discussed in next section while discussing estimators for these parameters. Whereas, 
    the adversary Eve can not infer parameters from these
    measurements without the knowledge of delays $\mbf{\Delta}$. 

\begin{figure}
  \centering
  \includegraphics[height=1.5in]{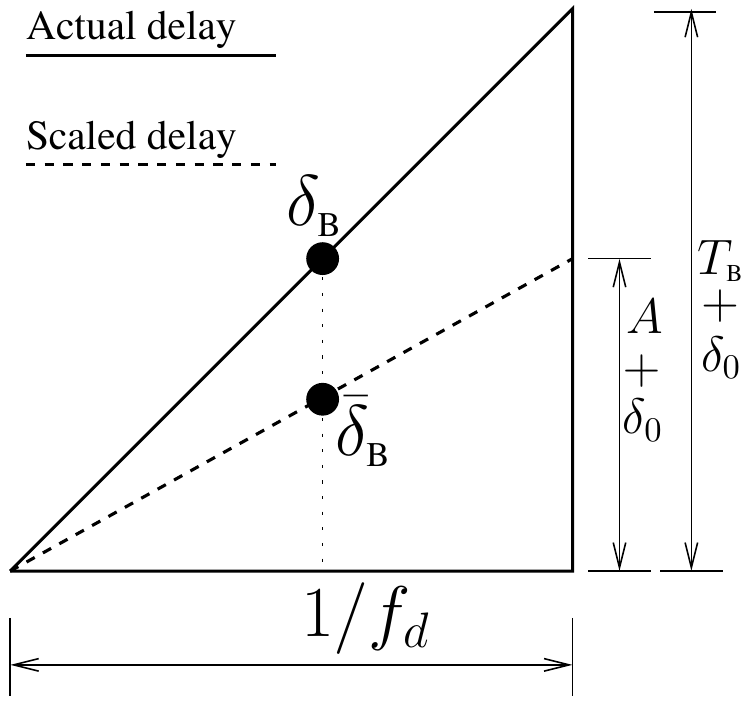}
  \caption{ Delay scaling at responding node Bob. The actual delay
    $\delta_{\text{\tiny B}}$ is mapped to the delay
    $\bar\delta_{\text{\tiny B}}$ to ensure the maximum height of
    sawtooth is '$A$'.}
  \label{TmA}
\end{figure}

%\subsubsection{Scaled \textsc{Respond} by Bob}
    As it can be seen from (\ref{eq:m33}), the outer modulus with respect to $T_{\text{\tiny B}}$ will have a peak-to-peak magnitude of
    $T_{\text{\tiny B}}$. The height  of the sawtooth reveals
    information on Bob's clock period in such measurements. As can be
    seen further in the discussion from (\ref{eveeq}) that Eve can
    estimate parameters of sawtooth from TDOA measurements while
    eavesdropping. Thus an eavesdropper to the RTT measurements
    can estimate Bob's clock period $T_{\text{\tiny B}}$, hence clock
    frequency $f_{\text{\tiny B}}$. To avoid this, as explained in
    section \ref{protocols}, time
    scaling of the delay by Bob is proposed before sending \textsc{Respond}.  As
    shown in the Fig.\ref{TmA}, delay $\delta_{\text{\tiny B}}$ with
    maximum value $T_{\text{\tiny B}}$ seconds is scaled to a delay
    $\bar\delta_{\text{\tiny B}}$ with maximum value $A$ seconds.

\begin{eqnarray}
  \bar\delta_{\text{\tiny B}} = \delta_{\text{\tiny B}}\frac{A +
  \delta_0}{T_{\text{\tiny B}} + \delta_0}.
  \label{delayscale}
\end{eqnarray}
$A$ is known to all. As shown in Fig.\ref{detailed_RTT}, Bob measures
$g(1)$ and computes $\delta_{\text{\tiny B}} = g(1) +
\delta_0$. Further, Bob computes $\bar\delta_{\text{\tiny B}}$ from
(\ref{delayscale}) and advances or delays transmitting
\textsc{Respond} by
$\delta_{\text{\tiny B}} \pm \bar\delta_{\text{\tiny B}}$ seconds. In
the shown examples of Fig.\ref{detailed_RTT} and Fig.\ref{TmA}, $A$ is
shown smaller than $T_{\text{\tiny B}}$. Hence Bob is shown advancing
the time of \textsc{Respond} transmission. With scaled
\textsc{Respond}, the sawtooth function (\ref{eq:m33}) in RTT protocol transforms with a
multiplicative factor $\frac{A}{T_{\text{\tiny B}}}$ as (\ref{hdef2}) in
the CLIMEX protocol. As a result, peak-to-peak amplitude of the sawtooth in
CLIMEX protocol is a
known to all value $A$. 
% \subsection{Measurements at own clocks}
% Further, Alice and Bob measure their own clock period and frequencies.
% \begin{equation}
%   y_{\text{\tiny A}}^{\text{\tiny T}} = T_{\text{\tiny A}} +
%   m_{\text{\tiny A }}, \ \ y_{\text{\tiny B}}^{\text{\tiny T}} = T_{\text{\tiny B}} +
%   m_{\text{\tiny B}}. \label{ownclock}
% \end{equation}
% Where, $m_{\text{\tiny A }}$ and $m_{\text{\tiny A }}$ are zero mean
% independent Gaussian noise sources with distributions
% $\mathcal{N}(0, 2\sigma_j^2)$.  After Alice, Bob takes turn in
% transmitting \textsc{Ping}s while adding random delays known to
% itself. Whereas, Alice responds after a scaled delay as discussed
% above. It is assumed that parameters of interest in the system are
% time invariant for the duration of collected measurements.

%Once Alice has collected its set of measurements, Bob initiates the same protocol and collects its own set of measurements $\mbf{y_{\text{\tiny B}}}|_{t''}$. 

\subsection{The noise model}\label{noisesec}There are two sources of noise in
CLIMEX protocol. One is jitter noise in clock edges of Alice and Bob
with assumed distribution $\mathcal{N}(0, \sigma^2 _j)$ and another is channel
noise from wireless transmission with distribution
$\mathcal{N}(0, \sigma^2 _c)$. In measurement models, (\ref{eq:m11}), (\ref{eq:m22}), (\ref{eq:m33}),
(\ref{climeq1}), (\ref{bob_eq1}) and (\ref{hdef2}), the noise vector
$\mathbf{n}$ consists of noises generated from from Alice's clock
noise while transmitting \textsc{Ping}, wireless channel and Bob's
clock noise in starting the delay generation subsequent to receiving
\textsc{Ping}. Hence,
$\mbox{cov}(\mathbf{n}) = (\sigma^2 _c + 2\sigma^2_j) \mathbf{I}
$. Similarly, the noise vector $\mathbf{w}$ consists of noises from
\textsc{Respond} generation and wireless channel.  Hence,
$\mbox{cov}(\mathbf{w}) = (\sigma^2_c + \sigma^2_j) \mathbf{I} $.

\begin{figure}
  \centering
  \includegraphics[height=2in]{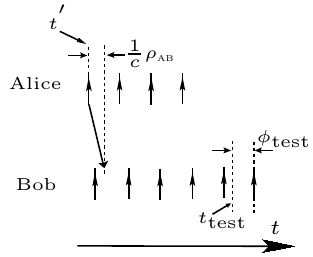}
  \caption{Alice predicting a phase $\phi_{\text{test}}$, at a
    predefined instance $t_{\text{test}}$ at Bob's end using estimates
    $\widehat{T}_{\text{\tiny B}},\, \widehat{\phi^{'}},$ and
    $ \widehat{\rho}_{\text{\tiny AB}}$ (\ref{phipredict}).}
  \label{phitest}
\end{figure}

\section{Estimation of Parameters by Alice and Bob}
\label{estim_AB} Alice and Bob  independently 
estimate parameters from the measurements obtained from CLIMEX
protocol. 
It is assumed that Alice and Bob know their own frequencies
$f_{\text{\tiny A}}$ and $f_{\text{\tiny B}}$. 
 Alice can estimate
parameters $f_d$, $\phi^{'}$ and $\rho_{\text{\tiny AB}}$ from the
measurements in (\ref{climeq1}) by minimizing a cost
function $J(f_d, \phi^{'}, \rho_{\text{\tiny AB}})$,
\begin{equation}
  \{\widehat{f_d}, \widehat\phi^{'}, \widehat\rho_{\text{\tiny AB}}\} = \arg \min J(f_d, \phi^{'},
  \rho_{\text{\tiny AB}}). 
\label{fd_phi_estim}
\end{equation}
The squared cost function at Alice can be given as
\begin{equation}
   J(f_d, \phi^{'}, \rho_{\text{\tiny AB}})\ \ \triangleq \ \ \left \|
    \mbf{y_{\text{\tiny A}}}|_{t'} \! -\!
    \mbf{h}(f_d,  \phi^{'}, \mathbf{\Delta}, \mbf{0})\!  -\! \delta_0 \mbf{1}  - \frac{2}{c}
    \rho_{\text{\tiny AB}} \mbf{1} \right \|^2.
  \label{eq:J_wls}
\end{equation}  
Above cost function can be minimized while estimating optimal value of
$\rho_{\text{\tiny AB}}$, 
\begin{equation}
  \widehat{\rho}_{\text{\tiny AB}}\  = \ \frac{ \mbf{y_{\text{\tiny A}}}|_{t'}  -
    \mbf{h}(f_d,  \phi^{'}, \mathbf{\Delta}, \mbf{0})\!  -\! \delta_0
    \mbf{1}}{2/c}. 
  \label{rhoeq}
\end{equation}
% Define
%$\mbf{r} = \mbf{y_{\text{\tiny A}}}|_{t'} \! -\!  \mbf{h}(f_d,
%\phi^{'}, \mathbf{\Delta}, \mbf{0})\!  -\! \delta_0$.
%\begin{equation}
%  \left[\widehat{f}_d, \widehat{\phi}^{'}\right] =  \argmin_{\tiny f_d\in F, \phi^{'} \in [0, 2\pi) }\left \| \mbf{r} - \frac{2}{c}\widehat{\rho}_{\text{\tiny AB}} \mbf{1} \right \|^2
%  \label{fd_phi_estim}
%\end{equation}
Because of double modulus nonlinearity of measurement model, we
propose estimating $\widehat{f}_d$ and $\widehat{\phi}^{'}$ by minimizing the
cost function as suggested in (\ref{fd_phi_estim}) by a grid search
over parameters \cite{satyam_clk1, kaybook}.  Now Alice can estimate Bob's clock
frequency $f_{\text{\tiny B}}$ from her estimate of $f_d$ i.e.,
\begin{eqnarray}
  \label{freqB@A}
  \widehat{f}_{\text{\tiny B}} &=& \widehat{f_d} - f_{\text{\tiny A}}.
                               \label{Alfb}
\end{eqnarray}
$\widehat{f}_{\text{\tiny B}}$ is Bob's clock frequency estimated by
Alice. Similarly, $f_{\text{\tiny A}}, \phi^{''},$ and
$ \rho_{\text{\tiny AB}}$ can be estimated by Bob, \textit{mutatis mutandis}.  

\subsection{Measuring the phase $\phi_{\text{test}}$ at a predefined
  instance $t_{\text{test}}$} Since Alice and Bob take turns in
collecting epochs, the phase estimates of the sawtooths, $\phi^{'}$
and $\phi^{''}$ will be different for both. Phase estimates
$\widehat\phi^{'}$ and $\widehat\phi^{''}$ at Alice and Bob are timestampped
by $t^{'}$ and $t^{''}$. Using each other's clock frequency and phase
estimate at a given time, they can estimate phases
of each other's clocks at any given time instance. Figure \ref{phitest} shows the phase
$\phi_{\text{test}}$ at Bob's clock which is the time measurement
between a pre-defined time instance $t_{\text{test}}$ and the
subsequent clock edge of Bob's clock. While Bob can measure
$\phi_{\text{test}}$ at its own clock, Alice can estimate it. The
pre-defined time $t_{\text{test}}$ can be provided by an external reference
clock. Or can be referenced at Bob with a specific received
\textsc{Ping} from Alice. Thus, they can establish a common phase
parameter among them.

Estimates $\widehat\phi^{'}$ and $\widehat\phi^{''}$
along with $\widehat{f}_{\text{\tiny A}}$ and $\widehat{f}_{\text{\tiny B}}$
helps Alice and Bob in predicting phases of each other's clock at any
instance of time. As can be seen in Fig.\ref{detailed_RTT}, Alice can
predict a phase $\phi_{\text{test}}$ at a preset time
$t_{\text{test}}$ on Bob's clock,
\begin{eqnarray}
  \widehat\phi_{\text{test}} = 2\pi - 2\pi\widehat{f}_{\text{\tiny B}}
  \!\!\!\!\mod_{\!\!1\!/\!\widehat{f}_{\text{\tiny B}}} \left ( t_{\text{test}} -
  t^{'} - \frac{\widehat{\phi^{'}}}{2\pi \widehat{f_d}} -
  \frac{\widehat{\rho}_{\text{\tiny AB}}}{c}  \right ).
  \label{phipredict}
\end{eqnarray}
Bob can also measure $\phi_{\text{test}}$ on its own clock. Thus
$\phi_{\text{test}}$  can be established as a shared parameter among
them. Thus from the CLIMEX protocol, Alice and Bob can share parameters $f_{\text{\tiny A}}$ , $f_{\text{\tiny B}}$, $\rho_{\text{\tiny AB}}$ and  $\phi_{\text{test}}$ among them.

\section{Secrecy Analysis}\label{secanal} There are many aspects of secure exchanges
using the CLIMEX protocol.  Robustness to several adversary models, non-observability of parameters to
adversary, secret bit generation from various independent parameters, deliberate dithering of measurements and
robustness to active adversaries. 
\label{sec_key}

\subsection{Deliberate dithering of measurements by introducing
$\mathbf{\Delta}$} 
As discussed previously, random added delays $\mbf{\Delta}$ to \textsc{Pings} dithers
the measurements for everyone. Alice and Eve knowing their respective
added $\mbf{\Delta}$ can retrieve parameters from measurements. This
is a very novel aspect of the CLIMEX protocol. While possible dynamic range of
estimates is quantified in terms of number of shared secret bits,
effect of adding $\mbf{\Delta}$  is not analyzed in this paper
quantitatively. However, it can be seen that arbitrary choice of
$\mbf{\Delta}$  can result in arbitrarily poor estimate of parameters
for an eavesdropper.  For every \textsc{Ping} transmitted there is a
$\Delta_i$ associated with it. Number of measurements in a set of
measurement by Alice or Bob can be in hundreds or thousands or even
more. The distribution of $\mbf{\Delta}$ can be arbitrary and depends
on the user. For a brute force attempt by Eve to guess $\mbf{\Delta}$,
Eve will have to attempt very large number of combinations of an
arbitrary distribution. Which can not even be attempted in lack of
knowledge of the distribution of $\mbf{\Delta}$.

\subsection{Eve as a passive  adversary, possible measurements and 
  estimation}
\label{adv_mod}
In this section  different adversary models will be discussed. The
adversary models will be discussed based on CLIMEX protocol discussed in 
previous sections. In this subsection, parameter estimation
capabilities of adversary from the measurements made on signals
received from Alice and Bob is explored, when Alice and Bob adhere to the CLIMEX
protocol.

Figure \ref{fig:ABE} shows Eve as an eavesdropper, a passive listener,
listening to message exchange between Alice and Bob. As shown in the
figure, $\rho_{\text{\tiny AB}}$ is the distance between Alice and
Bob, $\rho_{\text{\tiny AE}}$ is the distance between Alice and Eve
and $\rho_{\text{\tiny BE}}$ is the distance between Bob and Eve. 

As shown in Fig.\ref{1way2}, the third node
 Eve receives signals from Alice and Bob.  Eve is assumed to have at least the same capabilities and same
 hardware as Alice and Bob.  Eve's measurements are timestampped with
 time $t^{'''}$. Eve measurements shown as
 $\left [ y_{\text{\tiny E}}(1)\ y_{\text{\tiny E}}(2) \cdots \right
 ]^\top$  are time difference of
arrivals between receptions from Alice and Bob.

\subsubsection{Eve measuring time difference of arrivals of RTT
  measurements $\mbf{y_{\text{\tiny E}}}|_{t'} = [y_{\text{\tiny E}}(1) \: \cdots \:
y_{\text{\tiny E}}(N)]^\top $}
\begin{figure}
  \centering
   \includegraphics[width=3.2in]{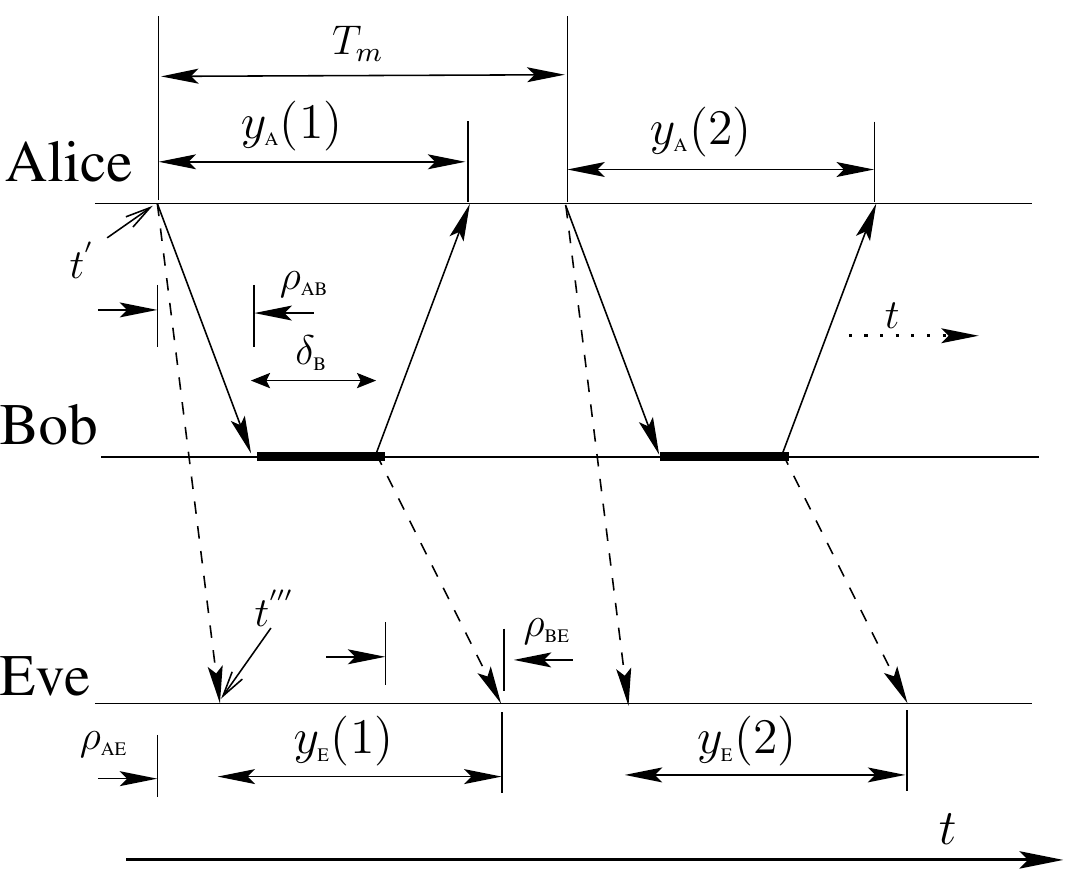}
   \caption{Conceptual timing diagram of one-way RTT protocol between
    Alice and Bob while Eve is a passive adversary. Eve is $\rho_{\text{\tiny AE}}$ and  $\rho_{\text{\tiny BE}}$ distances away from Alice and Bob. Eve collects RTT measurements $\left [y_{\text{\tiny E}}(1)\ y_{\text{\tiny E}}(2) \cdots \right
 ]^\top$.}
   \label{1way2}
 \end{figure}

From Fig.\ref{1way2}, measurements at Eve's end can be written as
\begin{equation}
  y_{\text{\tiny E}}(1) = \delta_{\text{B}}  + \rho_{\text{\tiny AB}} + 
  \rho_{\text{\tiny BE}} - \rho_{\text{\tiny AE}} + n_{\text{\tiny E}}.
  \label{ye1}
\end{equation}
Where $\delta_B$, the total delay at Bob's node given by
(\ref{eq:m22}).
% \begin{equation}
% \delta_B = \delta_0 + h(f_d,  f_{\text{\tiny B}}, \phi^{'''}, v).
% \end{equation}
Measurements in vector form at Eve can be written as
\begin{equation}
  % \mbf{y}|_{t' = \mbf{h}(f_d,\phi,\mbf{v}) + \delta_0 \mbf{1} +
  % \frac{2\rho_{\text{\tiny AB}}}{c} \mbf{1} + \mbf{C} +\mbf{n} \, ,
  \mbf{y_{\text{\tiny E}}}= \mbf{y_{\text{\tiny E}}}|_{t'''} = \boldsymbol{\delta}_{\text{B}}   + 
  r + \mbf{n_{\text{\tiny E}}} \, , 
\label{eveeq}
\end{equation}
where $t^{'''}$ is the timestamp when Eve starts recording the
measurements.  Sum of these distances,
$r = \rho_{\text{\tiny AB}} + \rho_{\text{\tiny BE}} -
\rho_{\text{\tiny AE}}$ contributes a constant term to the
measurements. The distances
$\rho_{\text{\tiny AB}}, \ \rho_{\text{\tiny BE}}$ and
$\rho_{\text{\tiny AE}}$ are non-observable to Eve in absence of
enough independent measurements.  As is evident from (\ref{eq:m11}),
(\ref{eq:m22}) and (\ref{eq:m33}), the vector
$\boldsymbol{\delta}_{\text{B}} $ contains all the information about
all the clock parameters in the RTT protocol. As shown in (\ref{eq:J_wls}) and  (\ref{rhoeq}), to estimate clock parameters Alice makes
use of $\boldsymbol{\Delta}$ which is known only to her. Since
$\boldsymbol{\Delta}$ is unknown to Eve hence Eve can not estimate
clock parameters when Alice and Eve collect measurement epoch from
CLIMEX protocol between them. 

However, if Alice and Bob collect RTT measurement epoch from the RTT
protocol between them as shown in Fig.\ref{detailed_RTT1}, Eve can
estimate clock parameters from time difference of arrival
measurements.  Eve can construct a new measurement vector
$\mathbf{\bar p}$ from the available measurements, which can be
defined as,
$\mathbf{\bar p}(f_d, T_{\text{\tiny B}}, \phi) =
\mathbf{y}_{\text{\tiny E}} - \mathbb{E}(\mathbf{y}_{\text{\tiny
    E}})$. Eve can further use the following cost
function, \begin{equation} J(f_d, T_{\text{\tiny B}}, \phi) \triangleq
  \left \| \mathbf{\bar p}(f_d, T_{\text{\tiny B}}, \phi) -
    \mbf{h}(f_d, T_{\text{\tiny B}}, \phi,\mbf{0})\right\|^2.
  \label{eq:J_wls2}
\end{equation}
The above cost function can be minimized to estimate $f_d$,
$T_{\text{\tiny B}}$ and $\phi$. Thus parameters can be estimated
under RTT protocol but not under the CLIMEX protocol.

\begin{figure*}
  \centering
  \begin{subfigure}{0.3\textwidth}
    \centering
    \includegraphics[height=1.7in]{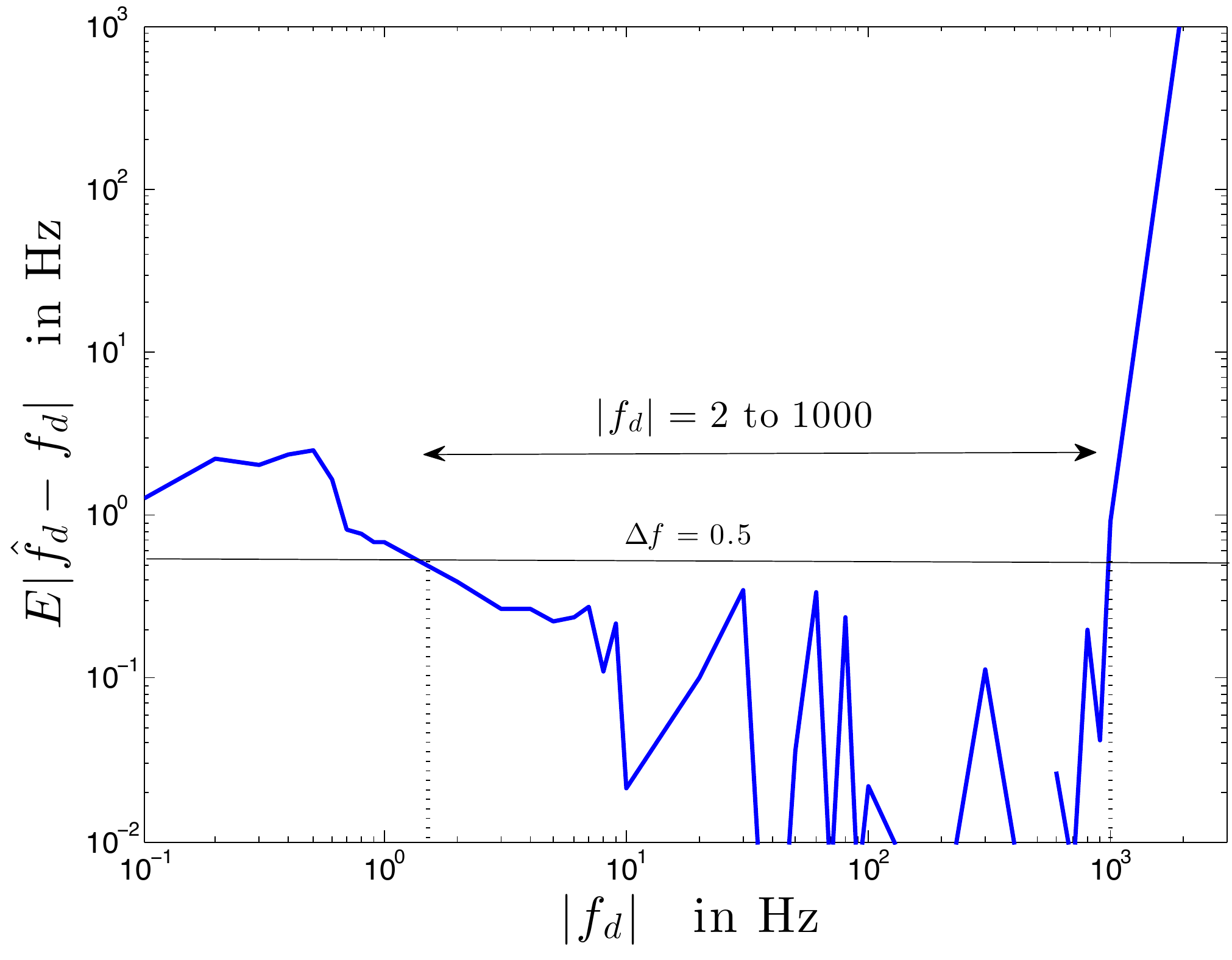}
    \caption{$\widehat f_d$ vs $f_d$ for low SNR}
    \label{fig:fdlimit1}
  \end{subfigure}
  \hfill
  \begin{subfigure}{0.3\textwidth}
    \centering
    \includegraphics[height=1.7in]{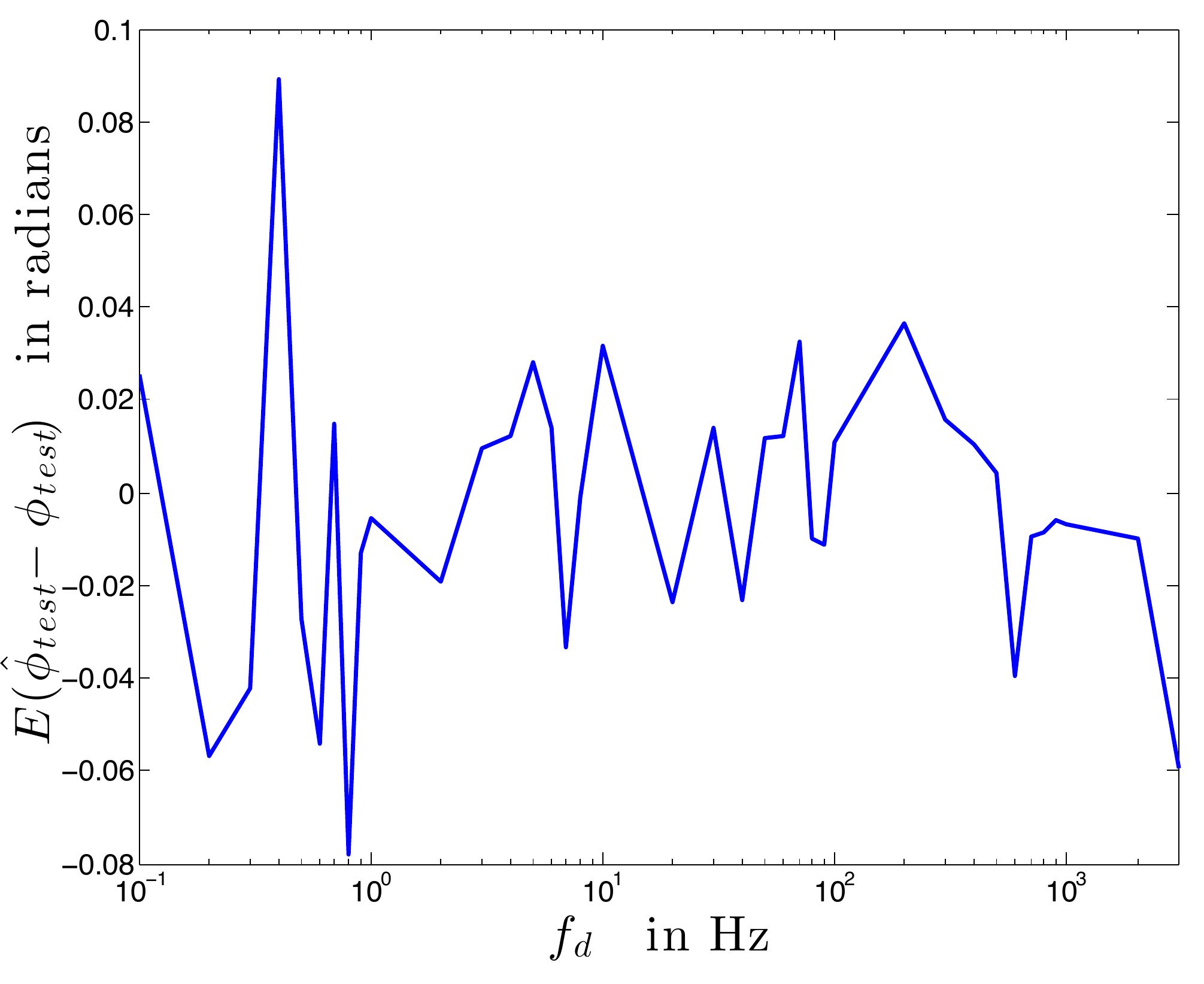}
    \caption{$\widehat\phi^{'}$ vs $f_d$ for low SNR}
    \label{fig:fdlimit2}
  \end{subfigure}
  \hfill
  \begin{subfigure}{0.3\textwidth}
    \centering
    \includegraphics[height=1.7in]{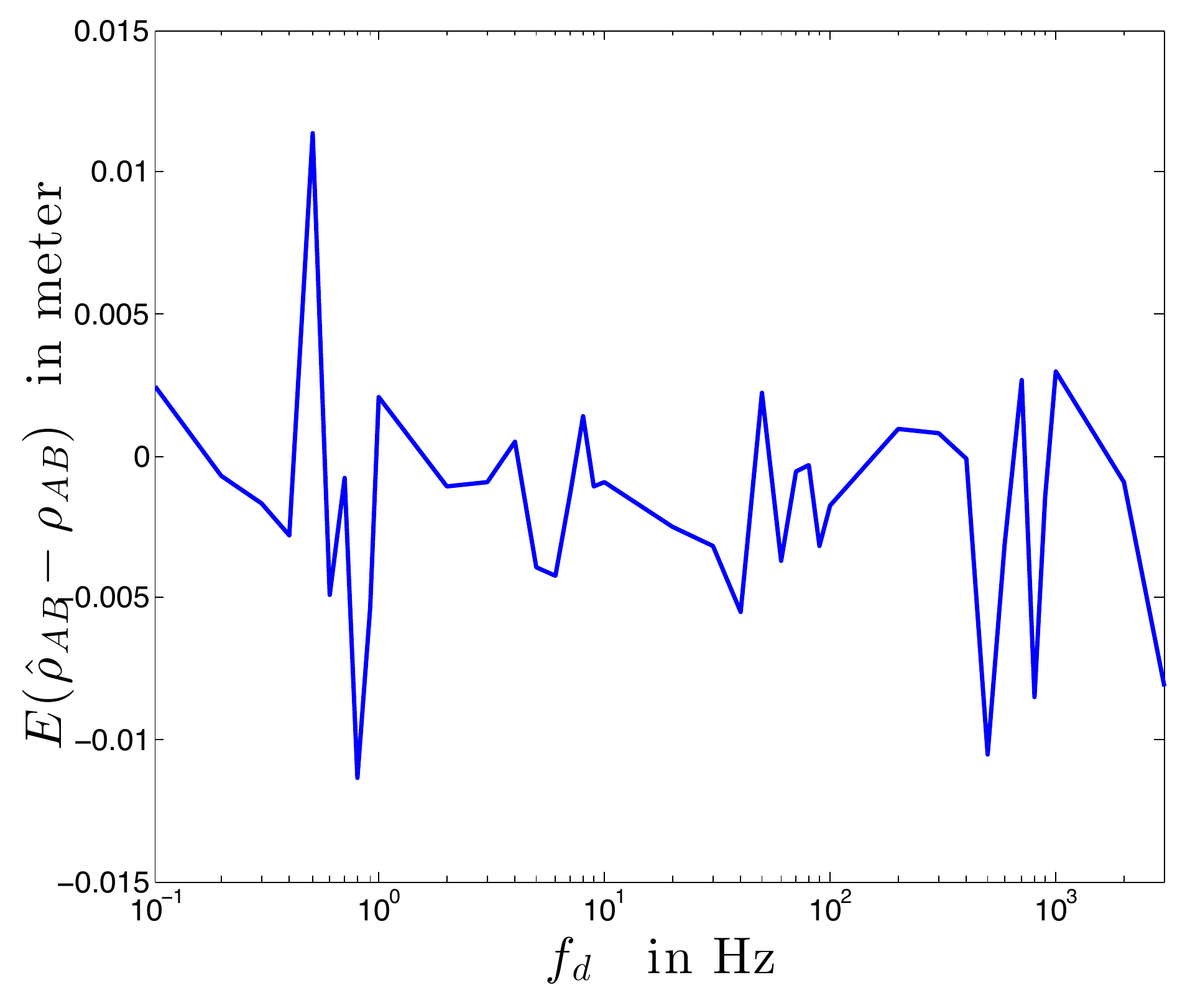}
    \caption{$\widehat\phi^{'}$ vs $f_d$ for low SNR}
    \label{fig:fdlimit3}
  \end{subfigure}
  \caption{$f_d$ estimation error with varying $f_d$. Simulated with
    $\sigma_j = 1 \text{ns}$, $\sigma_c = 2 \text{ns}$. Total duration of simulation is one
    second with $10^4$ 
    \textsc{Pings} per second. Change in $f_d$
    keeping other parameters of simulation same results in aliasing
    phenomenon, hence zig-zag fluctuations in simulated results. }
\label{fig:results}
\end{figure*}

\subsubsection{Eve measuring time difference of arrival of
  \textsc{Ping}s and its subsequent clock edge
}
\label{Eve_interarriv}
As can be seen in Fig.\ref{detailed_RTT1} that Bob can measure the
time between periodic arrivals from Alice and the next clock edge of
his own clock, $ h(1)$. Same measurement $g(1)$ is used by Bob in
CLIMEX protocol for delay scaling while responding as proposed while
discussing (\ref{delayscale}).  As discussed previously, epoch of such
measurements will result in sawtooth. From this it can be concluded that
measuring periodic time of arrival with a reference clock would give
rise to sawtooth having relative clock parameters of the transmitting
and the reference clock. Hence, Eve can measure time of arrival from
Alice or Bob with her own clock as a reference clock to estimate
$f_{\text{\tiny A}}$ or $ f_{\text{\tiny B}}$, assuming she would know
her own clock frequency.

However, in CLIMEX protocol the corresponding measurement would be
$g(1)$. As can be seen from (\ref{hdef2}), Eve would need to know
$\mathbf{\Delta}$ to make sense of such measurements. Hence, while
such measurements would reveal relative clock information in RTT
protocol but in CLIMEX protocol such measurements will not be of any
use to Eve in absence of knowledge of $\mathbf{\Delta}$. 

\subsection{Shared secret bit generation from  exchanged
  parameters} 
% \end{itemize}
% \subsection{{Ways of generating secure keys}}
The parameters estimated between Alice and Bob, $f_{\text{\tiny A}}$,
$ f_{\text{\tiny B}}$, $\phi_{\text{test}}$ and
$\rho_{\text{\tiny AB}}$ are all independent of each other. Alice and
Bob generate private keys using these estimated parameters.
% \begin{enumerate}[(1)]

\subsubsection{$\left(f_{\text{\tiny A}}, f_{\text{\tiny
          B}}\right)$} As discussed in section \ref{sys_mod}, clock
frequencies of Alice and Bob
$\left(f_{\text{\tiny A}}, f_{\text{\tiny B}}\right)$ can lie around
their nominal frequencies. The deviation around the nominal frequency
is usually specified in terms of PPM of the clock. However the nominal
frequency of the clock can be variable. Large differences in
$f_{\text{\tiny A}}$ and $ f_{\text{\tiny B}}$ and hence large values
of $|f_d|$ will require increased update rate (rate of sending \textsc{Pings}) of the system to
sufficiently sample the sawtooth waveform. Smaller values of $|f_d|$
would need longer measurement time. 
\begin{figure}
  \centering
  \includegraphics[height=2in]{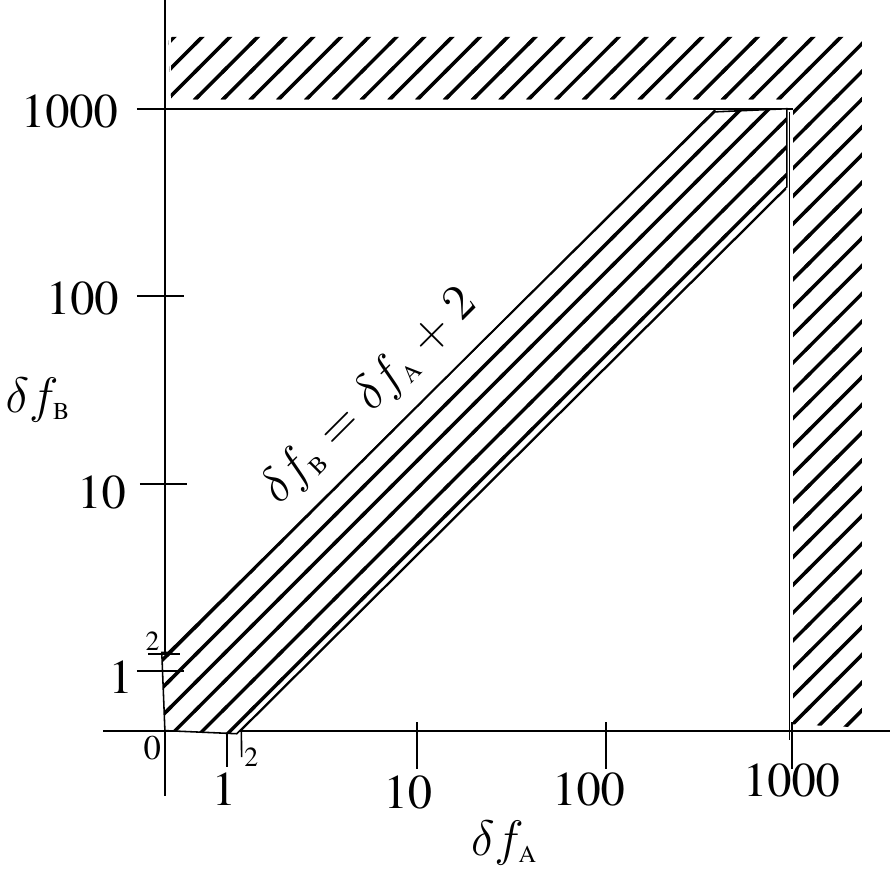}
  \caption{Valid combinations of $\delta f_{\text{\tiny A}}$ and
    $\delta f_{\text{\tiny B}}$ lie within un-shaded
    regions. Considering quantization bins of $1$\,Hz for
    $\delta f_{\text{\tiny A}}$ and $\delta f_{\text{\tiny B}}$, total
    number of combinations are area under the un-shaded region, equals
    $998^2$.}
  \label{fdcompute}
\end{figure}
To demonstrate 
number of possible secret bits from the measurements between Alice and
Bob, the experimental setup in \cite{satyam_clk1} is considered with
nominal clock frequencies at Alice and Bob, $f_0$ of $100$\,MHz, update
rate of $10^4$ \textsc{Pings} per second and clock deviation of
$10$\,PPM.  $f_{\text{\tiny A}} = f_0 + \delta\!f_{\text{\tiny A}}$,
$f_{\text{\tiny B}} = f_0 + \delta\!f_{\text{\tiny B}}$. Hence,
$|f_d| = |\delta\!f_{\text{\tiny A}} - \delta\!f_{\text{\tiny B}}|$
and
$\delta\!f_{\text{\tiny A}}, \delta\!f_{\text{\tiny B}} \sim
\mathcal{U}(-500, 500)$. However, the accuracy of estimating $\widehat{f}_d$
would determine the quantization of range of values of
$\delta\!f_{\text{\tiny A}}$ and $\delta\!f_{\text{\tiny B}}$.  Figure\ref{fig:results} is simulated using above parameters. Where deviation
in estimates are recorded for values of $f_d$. As can be seen from
Fig.\ref{fig:results}, for large values of $f_d$, undersampling of
sawtooth results in large errors. While for very low values of $f_d$,
period of sawtooth waveform is too large to be captured for simulation
duration. The non-monotonous, zig-zag  nature of curves in simulation
results is due to the aliasing effects of waveform sampling while
changing the sawtooth frequency.  

As
shown in Fig.\ref{fig:fdlimit1}, a limit is set on the performance of
the estimator as $E|\widehat{f}_d - f_d| \leq \Delta f$.  This sets frequency quantization bins of size $\Delta f$. Set of possible
combinations of $\!f_{\text{\tiny A}}$ and $\!f_{\text{\tiny B}}$ ,
$\mathcal{T}$, constraining the performance of estimating $f_d$ can be
written as
\begin{eqnarray}
  \mathcal{T}  :=  & \hfill \nonumber \\ 
&\!\!\!\!\!\!\!\!\!\!\!\!\!\!\!\!\!\!\left\{(\delta f_{\text{\tiny A}}, \delta f_{\text{\tiny B}}): 
    E|\widehat{f}_d - f_d| \leq \Delta f, \delta f_{\text{\tiny A}}\  \mbox{and}\  \delta f_{\text{\tiny B}} \in (1, 2,
  .. ) \right\}. \ \ 
\end{eqnarray}
Cardinality of the set $|\mathcal{T}|$ determines number of secret
bits extracted from uncertainty in values of $f_{\text{\tiny A}}$ and
$f_{\text{\tiny B}}$.  As shown in the Fig.\ref{fig:fdlimit1}, for
$\Delta f = 0.5$\,Hz, possible values of $f_d$ lies from $2$ to
$1000$\,Hz.  The number of combinations of $f_{\text{\tiny A}}$ and
$f_{\text{\tiny B}}$ pair can be computed as area under the un-shaded
regions in Fig.\ref{fdcompute}, when $\delta\!f_{\text{\tiny A}}$ and
$\delta\!f_{\text{\tiny B}}$ are quantized by $1$\,Hz bins. Area of
the un-shaded regions can be computed as sum of areas of the two
triangles in the Fig.\ref{fdcompute}, which is
$|\mathcal{T}|= 998^2$.  Thus, total number of secret bits extracted
from uncertainty of
$\left( f_{\text{\tiny A}}, f_{\text{\tiny B}}\right)$ combinations is
$N_f = \log_2(|\mathcal{T}|) \sim 20$\,bits.

\subsubsection{$\phi_{\text{test}}$} As can be seen in
Fig.\ref{fig:fdlimit2}, the parameter $\phi_{\text{test}}$ can be
resolved with precision of $0.1$ radian for the considered noise and system
parameters. So, the number of possible secret bits from uncertainties
in estimating $\phi_{\text{test}}$ is,
$N_{\phi} = \log_2(2\pi/0.1) \sim 6$\,bits.

\subsubsection{$\rho_{\text{\tiny AB}} $} From
Fig.\ref{fig:fdlimit3}, precision of estimating distance between Alice
and Bob can be upto $2$\,cm. Such a precision in range measurements can be
obtained even in present day commercial ranging devices
\cite{decawave}. Thus for a possible range of $100$\,meters between
Alice and Bob. Number of possible uncertainties and hence number of
secret bits are $N_\rho = 100/0.02 \sim 12$\,bits.

Thus, total number of possible secret bits considering the system and
noise parameters discussed above can be
\begin{equation}
  N = N_f + N_{\phi} + N_\rho \sim 38\,\mbox{bits}.
\end{equation}
Above analysis is simplified as the range of parameters and their bins are chosen conservatively.
However, it serves the purpose of demonstrating the possibility of extracting secret bits from the CLIMEX protocol.

\subsection{Eve an active adversary}
\label{sec:act1}
\begin{figure}
  \centering
  \includegraphics[height=3in]{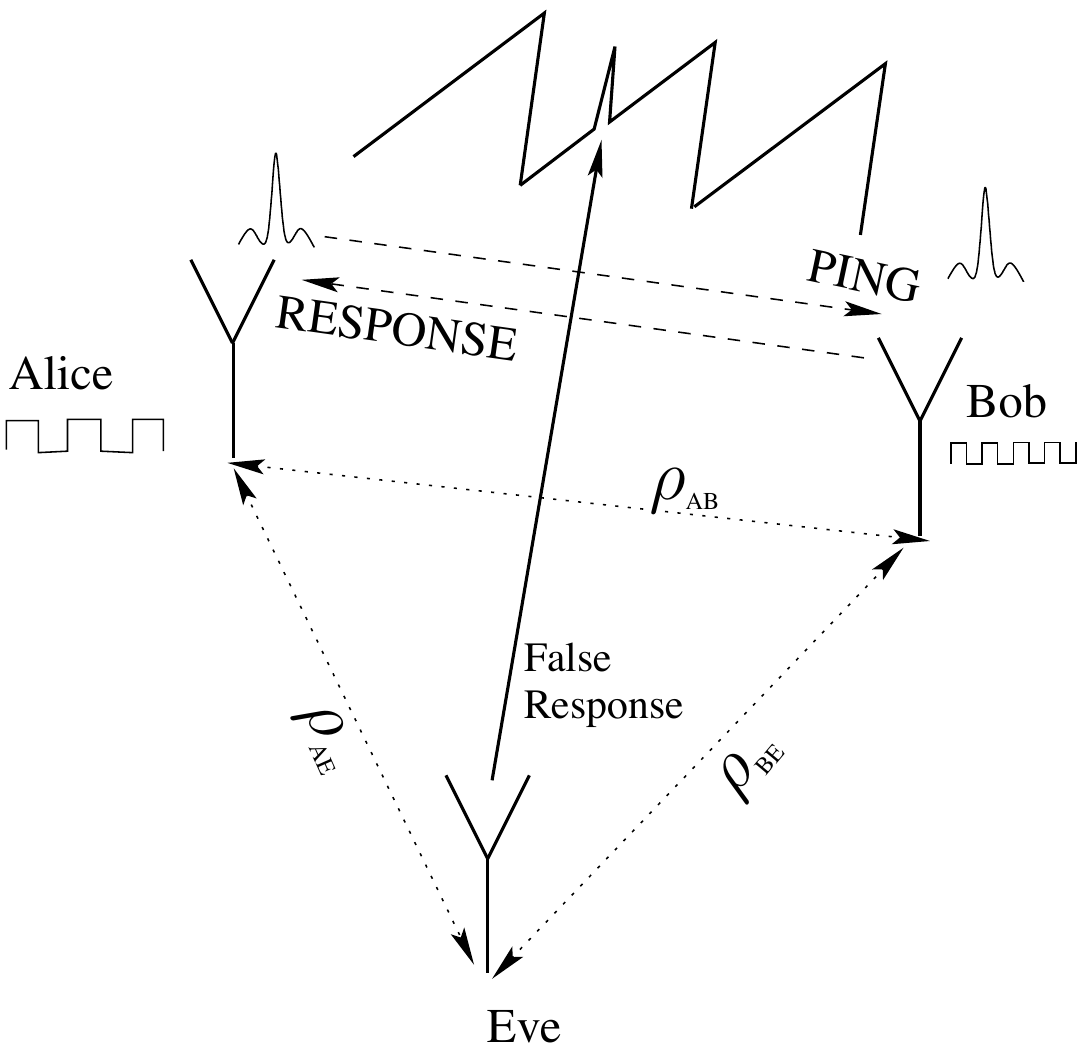}
  \caption{Eve trying to infiltrate the exchanges between Alice and
    Bob. Eve's response can show up as outlier in sawtooth measurement
    if Eve can not estimate all the distances in the setup besides
    estimating all clock parameters simultaneously. }
  \label{fig:infilterate}
\end{figure}

\begin{figure}
  \centering
  \includegraphics[height=2in]{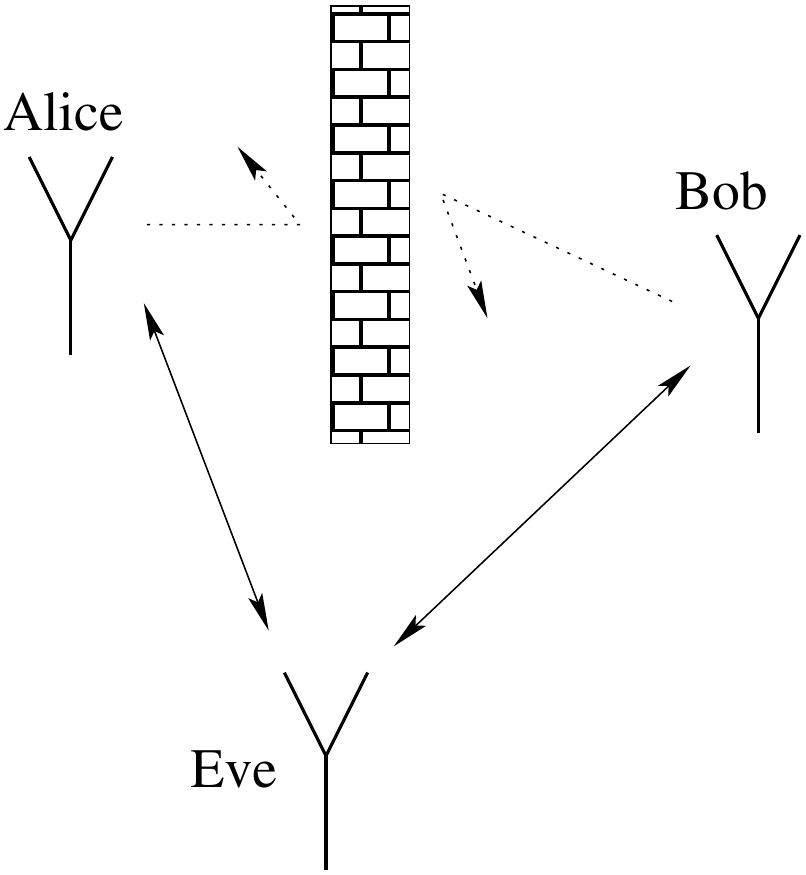}
  \caption{Alice and Bob are disconnected, Eve can talk to both and
    tries to impersonate Bob }
  \label{fig:impersonate}
%\vspace{-0.3in}
\end{figure}
Alice and Bob can track the generated
sawtooth waveforms from RTT measurements as valid signatures. 
For Eve to either impersonate or to infiltrate 
the exchanges without being detected, she has to ensure that the time of arrival of her signal at
Alice or Bob should fall on the sawtooth waveform being generated from the measurements.
 To achieve this, she  has to estimate all the distances,
$\rho_{\text{\tiny AB}}$,   $\rho_{\text{\tiny BE}}$ and 
$\rho_{\text{\tiny AE}}$ besides estimating all clock parameters for Eve to transmit at the
right time so to be able to fit its response on a sawtooth produced at
Alice or Bob.  Figure \ref{fig:infilterate} illustrates a response called '\textit{False Response}' 
from Eve attempting 
to infiltrate message exchange between Alice and Bob but instead showing up as an
outlier. It should be noted that Eve has to estimate more parameters
than Alice and Bob in order to infiltrate. Additionally, as
can be seen from (\ref{eveeq}), the three distances are non-observable
from the measurements between Alice and Bob. So, Eve does not get
enough information to infiltrate or impersonate the exchanges between
Alice and Bob. 

Eve as an active adversary can also try to participate
in mutual signal exchanges between Alice and Bob. Particularly in situations when direct
communication between Alice and Bob gets blocked without them knowing
it, as shown in Fig.\ref{fig:impersonate}. Here Eve gets an opportunity
to impersonate the responding node. In such situations, the sawtooth waveform
 in the measurement can come again to rescue. As discussed above, Eve
 will have to transmit at times with precise knowledge of all
 parameters. Even the sawtooth measurement would provide some
 robustness to jamming. Resistance
 against any timing attack through jamming or worm-hole attack can
 also be resisted while tracking the sawtooth signature in time or
 arrival of signals \cite{sec_posit_JSAC}. Correlation properties of sawtooth waveform
 would provide some level of resistance against jamming and similar attacks.

\subsection{Non-observable parameters} 
Clock frequencies of Alice and
Bob, $f_{\text{\tiny A}}$, $ f_{\text{\tiny B}}$ are non-observable to
Eve from measurements. As can be seen from the following arguments.

\begin{enumerate}
\item It is assumed that Alice and Bob know their own clock
  frequencies $f_{\text{\tiny A}}$ and $f_{\text{\tiny B}}$. 
\item Alice can estimate $f_d$ (\ref{fd_phi_estim}) and Alice can also trivially 
  estimate Bob's clock frequency $f_{\text{\tiny B}} = f_d - f_{\text{\tiny A}}$,
  (\ref{Alfb}).
\item Similarly, Bob can estimate its own and Alice's 
  frequencies.
\item It can be seen from (\ref{climeq1}),
(\ref{bob_eq1}) and (\ref{hdef2}) that $f_{\text{\tiny A}}$ and
  $f_{\text{\tiny B}}$, individual clock frequencies of
  Alice and Bob  are non-observable these RTT measurements. This is
  true even without using any random transmit delay $\mbf{\Delta}$.  
 It is assumed that  no one else knows Alice's and Bob's clock
 frequencies. In case of Alice and Bob not revealing it by themselves,
 it will require direct physical access to Alice's and Bob's clock to measure their
  frequencies.
\item As discussed in previous sections, Alice and Bob can estimate $f_{\text{\tiny A}}$ and
  $ f_{\text{\tiny B}}$. However, these frequencies are non-observable
  to Eve or anyone else.
  \item Usage of $\mbf{\Delta}$ for deliberate dithering of measurements further makes it difficult for Eve to eavesdrop. 
\end{enumerate}
Mathematically, it should be observed from (\ref{climeq1}),
(\ref{bob_eq1}) and (\ref{hdef2}), in RTT measurements collected
from CLIMEX protocol,  $f_{\text{\tiny A}}$ and
  $ f_{\text{\tiny B}}$ are non-observable and can not be
  estimated unless $f_{\text{\tiny A}}$ or
  $ f_{\text{\tiny B}}$ is known. It means that Eve can not estimate these parameters from any
possible measurements when Alice and Bob follow the CLIMEX protocol.

% \begin{itemize}

\section{Discussions}
\label{secdis}

The idea presented in this paper is conceived from the testbed
outcome. Clock parameters and distances are physical parameter which
can be controlled and can be setup to a good extent based on
necessity. Different parameters for generating secret bits discussed
above have different properties and relevance in the scheme. While
frequencies and the distance have unknown values in the system which
can be setup by the users in the system to certain precision, the
phase is usually out of control of the users. Frequency of the clock
and position of Alice or Bob can possibly be obtained by Eve by some other physical access or means. But
instantaneous relative phase of two clocks separated by a distance can only be obtained at the
time of operation. As discussed in previous section, the
parameter  $\phi_{\text{test}}$ amounts to nearly $6$ shared secret
bits in the considered setup, it is more infeasible to be revealed to
the adversary. 

In present day consumer electronic systems, range of parameters chosen
in previous section to derive number of bits can be larger. Update
rate can be possibly many orders higher. Distance estimation with
sufficient accuracy is possible of up to a kilo-meter. Clock frequency
of nodes can be in hundreds of mega-hertz. Overall, higher number of
shared secret bits is possible with aggressive selection of system
parameters. Number of shared secret bits can also be increased by successive execution of CLIMEX protocol.

Considering scope of future work, the proposed CLIMEX protocol can
also be applied for following general application scenarios
\begin{enumerate}
\item Secure ranging and secure
positioning in scenarios such as \cite{sec_posit_JSAC, secure_vehicle_Panos, 6510560, 7823100}.
\item Time synchronization in wireless networks can be secured with the CLIMEX implementation between nodes. While achieving the time synchronization at the same time in schemes such as \cite{zachariah2016scalable}.
\end{enumerate}

\section{Conclusion}
CLIMEX  as a secure wireless protocol is proposed for establishing secret bits between Alice and Bob, based on a round-trip-scheme where Alice and Bob sequentially set up the information exchange in a ping-response fashion.  A CLIMEX ping by Alice is securitized by an inherent synthetic clock edge or time-domain dithering, only known to Alice. The subsequent response by Bob is securitized by an amplitude-domain scaling. At Alice, a double-modulus nonlinear measurement model provides means for extracting the inherent physical parameters of both Alice and Bob, based on repeated measurements of Bob responses, and the unique knowledge of Alice’s own inherent physical parameters. The physical parameters of Alice and Bob form the foundation for the secret bits.  Subsequently, for reciprocity Bob is taking Eve's role, and \textit{vice versa}. By CLIMEX protocol, Bob is retrieving the secret bits, \textit{mutatis mutandis}.
For a passive adversary Eve with at least the same capabilities as Alice and Bob, it is shown that the physical parameters are unobservable. In addition, for an active impersonator Eve it is shown that its actions are detectable at Alice and Bob, manifested as outlier measurements.
Use of IEEE 802.15.4a like impulse radio UWB technology is considered for CLIMEX implementation. In-house testbed based on Xilinx Virtex-5 and ACAM time-to-digital-converters for required sub-clock resolution provided us with supporting experimental data and design parameters, showing potential for some 38 possible secret bits in typical use cases.  
Potential of CLIMEX can be foreseen for use cases with short to medium range distances (1-100 meters) between the nodes, including a variety of services for wireless communications, ranging and positioning.   

\bibliographystyle{IEEEtran} \bibliography{ref}
\end{document}